\documentstyle[12pt]{article}
\begin{document}

\def\ASCA{{\it ASCA\/}}
\def\ROSAT{{\it ROSAT\/}}
\def\Einstein{{\it Einstein\/}}
\def\Ginga{{\it Ginga\/}}

\title{\vspace{-1cm}
Discovery of the Central Excess Brightness in Hard X-rays 
in the Cluster of Galaxies Abell 1795} 

\author{
{\sc Haiguang} {\sc Xu}$^{1,2}$, {\sc Kazuo} {\sc Makishima}$^{1}$, 
{\sc Yasushi} {\sc Fukazawa}$^{1}$,\\
{\sc Yasushi} {\sc Ikebe}$^{3}$, {\sc Ken'ichi} {\sc Kikuchi}$^{4}$, 
{\sc Takaya} {\sc Ohashi}$^{4}$\\
{\small AND} {\sc Takayuki} {\sc Tamura}$^{1}$\\
{\small\it $^{1}$ Department of Physics, School of Science, University of Tokyo,} \\
{\small\it 7-3-1 Hongo, Bunkyo-ku, Tokyo 113, Japan}\\   
{\small\it $^{2}$ Institute for Space Astrophysics, Department of Physics, 
Shanghai Jiao Tong University,} \\
{\small\it Huashan Road 1954, Shanghai 200030, PRC} \\
{\small\it $^{3}$ Cosmic Radiation Laboratory, the Institute of Physical and Chemical Research,}\\
{\small\it 2-1 Hirosawa, Wako, Saitama 351-01, Japan}\\
{\small\it $^{4}$Department of Physics, Faculty of Science, Tokyo Metropolitan University,}\\
{\small\it 1-1 Minami-Ohsawa, Hachioji, Tokyo 192-03, Japan} \\
 }
\date{}
\maketitle
\vspace{1cm}
~~~~~~~~~~ Received  ~~~~~~~~~~~~~~~~~~~~~~~~~~~~~~~~~~~  ; ~~accepted
\clearpage
\begin{abstract}

Using the X-ray data from \ASCA, 
spectral and spatial properties of 
the intra-cluster medium (ICM) of the cD cluster Abell 1795 are studied,
up to a radial distance of $\sim 12'$ ($\sim 1.3$ $h_{50}^{-1}$ kpc). 
The ICM temperature and abundance are spatially rather constant, 
although the cool emission component is reconfirmed in the central region.
The azimuthally-averaged radial X-ray surface brightness profiles are 
very similar between soft (0.7--3 keV) and hard (3--10 keV) energy bands,
and neither can be fitted with a single-$\beta$ model
due to a strong data excess within $\sim5'$ of the cluster center.
In contrast, double-$\beta$ models can successfully reproduce the 
overall brightness profiles both in the soft and hard energy bands,
as well as that derived with the \ROSAT\ PSPC.
Properties of the central excess brightness are very similar 
over the 0.2--10 keV energy range spanned by \ROSAT\ and \ASCA.
Thus, the excess X-ray emission from the core region of this cluster
is confirmed for the first time in hard X-rays above 3 keV.
This indicates that the shape of the gravitational potential 
becomes deeper than the King-type one towards the cluster center.
Radial profiles of the total gravitating matter, 
calculated using the double-$\beta$ model, 
reveal an excess mass of $\sim 3 \times 10^{13}~ M_{\odot}$ 
within $\sim 150\ h^{-1}_{50}$ kpc of the cluster center. 
This suggests a hierarchy in the gravitational potential
corresponding to the cD galaxy and the entire cluster. 
\end{abstract}

\noindent
{\it subject headings}: galaxies: clusters: individual (Abell 1795) -- galaxies: 
elliptical and lenticular, cD -- X-rays: galaxies

\clearpage
\section{INTRODUCTION}

Imaging soft X-ray observations have established 
that nearly two thirds of clusters of galaxies,
particularly those with cD galaxies, 
show ``central excess X-ray emission''
(Jones \& Forman 1984; Edge \& Stewart 1991ab; Edge, Stewart, \& Fabian 1992).
That is, diffuse X-ray emission from their intracluster medium (ICM) 
exhibits centrally peaked surface brightness profiles
when compared with $\beta$ models (e.g., Sarazin 1988) 
which describe an isothermal gas sphere 
hydrostatically confined in a King-type potential (King 1972).
Since the X-ray volume emissivity from the ICM is 
proportional to the ICM density squared,
the excess X-ray emission generally implies 
an excess ICM density in the cluster core region.
This may in turn result from two apparently independent
(and mutually non-exclusive) mechanisms.
One is decrease in the average ICM temperature 
(or presence of a cooler ICM phase) in the cluster core region, 
particularly due to cooling flows (see Fabian 1994 for a review);
then the $\beta$-model is no longer valid since the gas is not isothermal,
and the central ICM density increases so as to maintain the pressure.
The other is deviation of the gravitational potential
from King-type ones, in such a way 
that the potential exhibits an additional drop in the cluster center
which attracts an excess amount of ICM.

Among these two mechanisms, the former
effect is usually thought to be significant, 
because the ICM is often multi-phased in the central regions of cD 
clusters (Fabian et al. 1994; Fukazawa et al. 1994; Matsumoto et al. 1996), 
and the radiative cooling time of the ICM in such regions is estimated 
to be generally shorter than the Hubble time (Fabian 1994). 
In contrast, the effect of potential shapes on the central X-ray excess 
has so far remained inconclusive, because the imaging X-ray telescopes 
were available only in energy ranges below $\sim 3$ keV 
where the X-ray brightness is usually more sensitive to the 
temperature decreases than to the detailed potential shapes.

With \ASCA\ (Tanaka, Inoue \& Holt 1994), the energy range 
available for X-ray imaging studies has been expanded up to 10 keV,
and the spectroscopic capabilities have been drastically improved.
Observations of the Fornax cluster with \ASCA\ have revealed 
that the strong central excess brightness around its cD galaxy, 
NGC~1399, is a consequence of the shape of the gravitational potential,
which requires two distinct spatial scales corresponding to 
the entire cluster and the central galaxy 
(Makishima 1995, 1996; Ikebe 1995; Ikebe et al. 1996).
The optical surface photometry of NGC~1399 shows 
essentially the same effect (Schombert 1986).  
Similarly, the prominent central excess emission from the Hydra-A cluster,
previously known in soft X-rays (David et al.1990),
has been detected in harder X-rays up to 10 keV (Ikebe et al. 1997).
Therefore the potential of this cluster is inferred to exhibit 
a marked central dimple to confine the excess hot ICM.
Even in cD clusters with prominent central cool emission, such as
the Centaurus cluster (Fukazawa et al. 1994; Ikebe 1995)
and the Virgo cluster (Matsumoto et al. 1996),
non-cooled ICM was found to fill a major fraction of the cluster core volume;
two-phase modeling of the ICM in the Centaurus cluster has yielded a 
potential profile with a noticeable feature in the central $\sim 50$ kpc (Ikebe 1995).

Similar effects have been observed from smaller systems as well.
The giant elliptical galaxy NGC~4636 was revealed to sit at the center 
of two concentric X-ray brightness components of similar temperatures, 
having distinct scales of $\sim 30$ kpc and $> 200$ kpc
(Matsushita 1997; Matsushita et al. 1996, 1997). 
Therefore NGC~4636 must be surrounded by 
hierarchically nested two gravitating halos, 
one corresponding to the galaxy itself and the other to a galaxy group
which is otherwise difficult to recognize.
The nested two-component X-ray brightness distributions have also
been  detected with {\it ROSAT} from several galaxy groups 
(Mulchaey \& Zabludoff 1997).

Thus, recent X-ray observations suggest
that the gravitational potential shape has a significant effect 
on the phenomenon of central excess X-ray emission.
It may be that the cluster potential generally consists
of doubly-nested hierarchical components, 
corresponding to the cluster and the cD galaxy.
Alternatively, the potential shape of a self-gravitating system 
may normally exhibit a significant central cusp,
as suggested by some gravitational lensing measurements  
(e.g., Wu \& Hammer 1993; Miralda-Escude \& Babul 1995)
and $N$-body simulations (Navarro, Frenk, \& White 1996).
If either (or both) of these possibilities are true, 
the central excess brightness of clusters should
be detectable in hard X-rays as a rule rather than the exception.
In this paper, we examine this possibility using the prototypical
central-excess cluster, Abell 1795 (hereafter A1795). 

A1795, a richness class 2 cluster at a redshift of $z=0.0616$, is one of the 
most equilibrated clusters with nearly circular X-ray isophotes (Briel \& Henry 1996). 
The soft X-ray brightness within $4'-5'$ of its cD galaxy (MCG +05-33-005)
exhibits a strong excess, above a $\beta$-model fitted to the 
outer-region brightness profile (Jones \& Forman 1984; Briel \& Henry 1996).
In the 0.5--3 keV band, the central excess emission is so strong 
that it accounts for 30--50\% of the total luminosity in the same energy band.
Mainly based on these morphological results in soft X-rays, 
the cooling flow rate in this cluster was estimated to be as large as 
$\dot{M} =478 \ h_{50}^{-2} ~ M_{\odot}$ yr$^{-1}$ with {\it EXOSAT} 
(Edge, Stewart, \& Fabian 1992),
or $\dot{M}=512 \ h_{50}^{-2} ~ M_{\odot}$ yr$^{-1}$ with \ROSAT\ (Briel \& Henry). 
Spectroscopically, the ICM of A1795 has an 
emission-weighted mean temperature of $5.3$ keV
(Hatsukade 1990; Tsuru 1992), together with a central cool emission component
(Fabian et al. 1994; Mushotzky et al. 1995;  Briel \& Henry 1996).   
The spectroscopy with \ASCA\ (Fabian et al. 1994) provided a somewhat 
lower cooling-flow rate of $\sim 131 \ h_{50}^{-2} ~ M_{\odot}$ yr$^{-1}$.
Outer regions of A1795 are reported to be rather 
isothermal (Mushotzky 1994; Ohashi et al. 1997).
Besides all these X-ray studies, 
an optical surface photometry of A1795 (Johnstone, Naylor, \& Fabian 1991)
also illustrates an elongated large-scale halo around the cD galaxy,
which is similar to that of the Fornax cluster (Schombert 1986). 
Clearly, this object is ideal to our search 
for the hard X-ray excess in the central region.

Using the \ASCA\ data of A1795, we have actually detected the central 
excess brightness in the hard energy band up to 10 keV, and found
that the excess percentage in the 3--10 keV band
is very similar to those found below $\sim 3$ keV.
We also analyzed the archival \ROSAT\ data 
and obtained a consistent result. 
Unless stated otherwise, 
we use the $90$\% confidence level throughout this paper,
and assume the Hubble constant to be $H_{0}=50$ $h_{50}$ km s$^{-1}$ Mpc$^{-1}$. 
We also employ the solar abundance ratios from Anders \& Grevesse (1989), 
with Fe/H$=4.68\times10^{-5}$ by number.

\section{OBSERVATION}

\ASCA\ observations of A1795 were conducted for $\sim 40$ ks 
on 1993 June 16-17, during the PV (Performance Verification) phase.
The field of view was centered at
$\alpha =13^{\rm h}48^{\rm m}30^{\rm s}.4$,  
$\delta =26^{\circ} 37' 39''.7$ (J2000).
The data from the GIS (Gas Imaging Spectrometer; Ohashi et al. 1996; 
Makishima et al. 1996) were taken in the normal PH mode,  
and those of the SIS (Solid-state Imaging Spectrometer; Burke et al. 1994) 
were acquired in 4 CCD Bright/Faint mode. 
The same dataset was already utilized by Fabian et al. (1994),
Mushotzky (1994), Mushotzky et al. (1995), and Ohashi et al. (1997).

We screened the GIS and SIS events with the standard \ASCA\ data selection criteria.
Specifically, we required the telescope view direction above the Earth's limb 
to be $> 10^\circ$ for the GIS data, $> 10^\circ$ for the night-time SIS data,
and $> 20^\circ$ for the day-time SIS data to avoid the light leakage.
The minimum cutoff rigidity was set to be 8.0 GeV c$^{-1}$ 
for both the GIS and the SIS. These screening processes have provided $\sim 35.8$ ks
and $\sim 37.3$ ks of good exposures with the GIS and the SIS, respectively.

Figure 1 shows GIS images of A1795 in the 0.7--3 keV and 3--10 keV bands,
where the cosmic X-ray background and internal detector background have been 
subtracted by utilizing observations onto several blank sky fields.
The two images are very similar to each other, both having a good circular symmetry. 
The X-ray brightness centroid is found at 
$\alpha =13^{\rm h}48^{\rm m}48^{\rm s}.9$ and  
$\delta =26^{\circ} 36' 20''.4$ (J2000),
with a typical uncertainty of $1'$.
This agrees with the optical position of MCG +05-33-005 within $\simeq1.1'$.
The X-ray emission extends at least to $20'$:
considering that $1'$ corresponds to $\simeq108~h^{-1}_{50}$ kpc,
the ICM emission is thus detectable at least up to $2.2~h^{-1}_{50}$ Mpc.

A Seyfert 1 galaxy, EXO 1346.2+2645, 
is located about $6'$ south of the A1795 center
and is included in the field of view of both the GIS and SIS.
Since the X-ray flux of this object is about 6\% of that of A1795,
we exclude a region around it in the following data analysis.
There is yet another active galactic nucleus (AGN) to $\sim 14'$ south of A1795,
emitting a flux which is 2\% of A1795.
We accordingly limit our analysis to within $12'$ from the cluster centroid.
Although the extended point spread function (PSF) of the \ASCA\ 
X-ray telescope (XRT; Serlemitsos et al. 1995; Tsusaka et al. 1995)
causes some fraction of X-rays from these AGNs to leak into the analysis region,
the effect is confirmed to be negligible 
compared with the photon counting errors.

\section{SPECTRAL ANALYSIS}

\subsection{Extraction of the Spectra}

Using both the GIS and SIS, we accumulated photons in five annular regions 
of $r<2'$, $2'<r<4'$, $4' < r <6'$, $6'< r < 8' $, and $8'<r<12'$,
where $r$ is the projected radius from the X-ray brightness centroid. 
However, due to the smaller field of view of the SIS ($\sim 22' \times 22'$) 
and an offset of the cluster centroid from the detector center (Fig.1), 
the SIS data only partially covers the regions outside $r=6'$.
The background was subtracted from each spectrum, 
in a similar way as mentioned in \S2.
Corresponding spectra from the two GIS detectors (GIS2 and GIS3) 
were then combined together; 
so were those from the two SIS detectors (SIS0 and SIS1). 
We show these spectra in Figure 2, 
where we combined four outer-region spectra for presentation.

As seen in Figure 2, all these spectra exhibit strong Fe-K emission lines.
In addition, both instruments reveal somewhat enhanced emission 
at $\sim 1$ keV in the spectra from the inner region ($r<2'$),
which is indicative of an additional cool component 
as already reported by Fabian et al. (1994).  
However in the outer regions ($r>4'$), 
the spectral shapes do not depend significantly on the radius.
Thus, the ICM of this cluster has relatively uniform properties in outer regions, 
in agreement with reports by Mushotzky (1994) and Ohashi et al. (1997),
who studied temperature and abundance distributions using the same \ASCA\ data.

\subsection{Conventional Spectral Fitting}
It is well known that the extended PSF of the \ASCA\ XRT 
causes difficulties in the spectral analysis of extended sources
(e.g. Takahashi et al. 1995; Markevitch et al. 1996; 
Ikebe et al. 1997; Ohashi et al. 1997).
In the case of clusters of galaxies,
X-rays from bright central regions are scattered by the PSF 
into fainter peripheral regions of the target, 
thus contaminating the outer-region spectra.
Furthermore, the scattered photons form a harder spectrum 
than their original one, 
because the PSF has more prominent outskirts in higher energies.
This often produces an artificial outward temperature increase.

In spite of these problems, we first performed a brief spectral analysis 
in a conventional way just to grasp gross properties of the X-ray emission. 
That is, we jointly fitted the SIS and GIS spectra at each radius
with the single temperature Raymond-Smith emission model (Raymond \& Smith 1977),
while the PSF effects are not taken into account.
The absorption column density and the heavy-element abundances were left free, 
whereas the abundance ratios were constrained to obey the solar ratios. 

This conventional analysis yielded mostly acceptable fits to the spectra,
and the temperature turned out to be in the range 5 -- 8 keV
with a mild outward increase due to the instrumental artifact (Takahashi et al. 1995).
Also there is a slight temperature drop at the center, 
due probably to the cool component (Fabian et al. 1994; Briel \& Henry 1996).
Except these, the cluster appears to be rather isothermal in agreement 
with Mushotzky (1994) and Ohashi et al. (1997) who used the same \ASCA\ dataset.
The metallicity is consistent with being radially constant at $\sim 0.3$ solar.
There is some hint of excess absorption in the inner regions 
up to a few times $10^{20}$ cm$^{-2}$,
above the Galactic line-of-sight value of $1.1\times10^{20}$ cm$^{-2}$, 
but the effect is not as significant as in some other clusters (White et al. 1991).

>From these fits, we estimate the X-ray luminosity of A1795 to be 
$(10.0 \pm 0.1) \times10^{44}~ h_{50}^{-2}$ erg s$^{-1}$ in 0.5--3 keV or
$(10.4 \pm 0.1) \times10^{44}~ h_{50}^{-2}$ erg s$^{-1}$ in 2--10 keV,
both integrated up to $r=12'$ ($1.3~h^{-1}_{50}$ Mpc) and corrected for absorption. 
The former value is close to the 0.5--3 keV 
luminosity of $(7.25 \pm 0.07) \times10^{44}$ erg s$^{-1}$
obtained with \Einstein\ within $r=12'$ (Jones \& Forman 1984).
Similarly, our 2--10 keV luminosity is in a good agreement with the value 
of $1.1\times10^{45}$ erg s$^{-1}$  (Hatsukade 1989), 
obtained with the \Ginga\ LAC in the same 2--10 keV range
but over a much wider field of view of $1^\circ \times 2^\circ$. 
Therefore, contribution to the X-ray luminosity from regions 
outside $r \sim 12'$ is inferred to be negligible.
Note that these \Einstein\ and \Ginga\ measurements both refer to $h_{50}=1$,
and that these \ASCA\ luminosities are approximately free from the PSF effects.

\subsection{Comparison with a Simulated Isothermal Cluster}
In order to derive the ICM temperature distributions considering the PSF effects,
we must solve the spectral couplings among different regions 
on the focal plane (e.g. Markevitch et al. 1996; Ikebe et al. 1997).
However such a detailed spectral investigation is beyond the scope of this paper.
Instead, we examine the \ASCA\ spectra against the hypothesis
of constant ICM temperature, taking the PSF effects fully into account.

For this purpose, we performed a Monte-Carlo simulation 
(Takahashi et al. 1995; Ikebe 1995)
which converts a model cluster into the simulated \ASCA\ data.
That is, we generate a number of Monte-Carlo photons 
by specifying their spatial and spectral probability distributions.
The photons are then ``reflected'' by the XRT and ``detected'' either 
by the GIS or the SIS, according to the known instrumental responses.
This tells us how a ``known'' cluster looks like when observed with \ASCA.
The XRT+GIS angular response is represented by actual images of Cygnus X-1
(a bright Galactic X-ray binary) obtained with the GIS (Takahashi et al. 1995).
However Cygnus X-1 is too bright for the SIS, 
so that this Monte-Carlo method cannot be applied at present to the SIS data.

In the simulation, we employ the Raymond-Smith model to express 
the spectral probability distribution of Monte-Carlo photons.
We assume the model cluster to have a free but spatially constant temperature,
and fix the model cluster abundance at $0.30$ solar
according to the results of \S 3.2.
The spatial probability distribution of Monte-Carlo photons 
is modeled by a $\beta$-model,
which describes the radial X-ray surface brightness profile as
\begin{equation} 
   S(r)=S(0) \left[ 1+ \left( \frac{r}{a} \right)^2 \right]^{-3\beta+ \frac{1}{2}} 
\end{equation}
where $S(0)$ is the central brightness, 
$r$ is again the projected radius, $a$ is the core radius, 
and $\beta$ is so-called $\beta$ parameter.    
We chose $a=0'.8$ and $\beta=0.56$ for our simulation. 
Later we in fact find that the brightness profiles of 
A1795 cannot be described by a single-$\beta$ model.
However, the $\beta$-model employed here can describe 
the \ASCA\ and \ROSAT\ brightness profiles of A1795 to within 5\% as we show later. 
The difference between this approximation and the true brightness distribution
produces negligible effects in the simulated spectra.

We generated about $10^6$ simulated GIS events,
which were subsequently sorted into 5 annular-region spectra 
exactly in the same way as the actual data.
Then, at each radius range, we directly compared the actual and simulated GIS spectra,
by subtracting the latter (after an appropriate re-scaling) from the former
in channel-by-channel way, and evaluating residuals via the chi-square tests.
The re-scaling factor was adjusted at each radius to minimize the chi-square.
We then varied the model cluster temperature, to optimize the overall fit goodness.
The model cluster with a temperature of $6.2$ keV 
has been found to best reproduced the GIS spectra,
yielding $\chi^2/\nu$ = 82/106, 92/106, 65/106, 53/106 and 82/106,
from the inner to outer annular regions.
Therefore we conclude that the observed GIS spectra are consistent with 
those expected from an isothermal cluster at a temperature of 6.2 keV,
when the PSF effects are fully taken into account.
This ``best-fit'' temperature is estimated to be uncertain by $\sim 0.3$ keV.

Figure 3 shows the ratio (instead of difference) 
between the actual and simulated GIS spectra,
in the innermost and $6'< r < 8'$ radius ranges.
As expected from the above results, 
the ratios are approximately constant across the energy range.

\subsection{Central Cool Component}

Although the overall cluster is nearly isothermal as confirmed above,
the central cool component detected by Fabian et al. (1994) reveals itself 
in the form of enhanced Fe-L complex (at $\sim 1$ keV) at about 1 keV,
in the innermost spectra of Figure 2 and the spectral ratios of Figure 3a.
To investigate this cool component, 
we extracted the SIS and GIS spectra from the inner region of $r<3'$.
This region, having by far the highest brightness, 
is little ($< 5$\%) affected by photons 
which are scattered from outer regions by the PSF into this region.
(However, outer regions are strongly affected by photons from the central region.)
We are therefore justified to analyze the spectra using ordinary energy responses
calculated at the position of the cluster centroid, 
without considering X-rays from outer regions due to the PSF effects.

We tried to fit these spectra jointly with a Raymond-Smith model
whose temperature is fixed at 6.2 keV, according to the result of \S 3.3.
However as shown in Table 1, the fit was totally unsuccessful 
even when the absorbing column density 
and the overall metallicity were allowed to float.
We then let the temperature to vary freely.
This gave a much improved fit together with a somewhat lower temperature (Table 1),
which is close to that measured with {\it Ginga} (5.3 keV),
but the fit still remained unacceptable.
We therefore introduced another Raymond-Smith component of lower temperature,
which is constrained to have the same abundance and the same absorption 
as the first component.
This two-temperature (2T) fit has been mostly successful as shown in Table 1.
The hot-component temperature ($6.54^{+0.56}_{-0.50}$ keV),
thus derived jointly with the two instruments,
agrees with that obtained via the simulation in \S3.3.
We also tried another two-temperature modeling after Fabian et al. (1994), 
who assumed that the hot component is absorbed with the 
Galactic line-of-sight column 
whereas the cool component is further absorbed by an excess column density. 
This model gave as good a fit ($\chi^{2}/\nu=380/275$) to the data,
and the derived results are essentially the same as obtained above.

>From the two-temperature fit,
we calculate the 0.5--3 keV luminosity of the cool component to be 
$L_{\rm cool} = (1.43\pm 0.39)\times10^{44}~ h_{50}^{-2}$ erg s$^{-1}$,
after removing absorption. 
This is $14\pm 4$\% of the total 0.5--3 keV luminosity described in \S 3.2.
The alternative modeling used by Fabian et al. (1994) yields a consistent value;
$L_{\rm cool} = (1.62 \pm 0.46) \times 10^{44}~ h_{50}^{-2}$ erg s$^{-1}$
also after removing absorption.
When these values of $L_{\rm cool}$ are substituted into equation (2) 
of the review by Fabian (1994), 
we get mass deposition rates of $\sim 100~ M_{\odot}$ yr$^{-1}$,
which are consistent with those obtained via the direct spectral fitting 
in terms of the cooling-flow model.
We cannot however compare our cool-component luminosities
with that derived by Fabian et al. (1994), 
which is not given explicitly in their publication.

In order to more directly confirm the consistency 
with the result of Fabian et al. (1994),
we tried an isothermal model plus an absorbed cooling flow model 
on       the central ($r<2'$) SIS0 spectrum.  
The mass deposition rate turned out to be 
$138^{+82}_{-52} \ h_{50}^{-2} ~ M_{\odot}$ yr$^{-1}$, 
with $\chi^{2}/\nu =306/230$.
This reconfirms the report by Fabian et al. (1994)
who used the same SIS data.

\section{RADIAL BRIGHTNESS PROFILES}

\subsection{X-ray Brightness Profiles}

In Figure 4, we show radial count-rate profiles of A1795
in two representative energy bands,
without correction for the XRT vignetting or the XRT PSF effects. 
They were derived by azimuthally averaging the counts 
from GIS2+GIS3 and SIS0 separately around the X-ray centroid, 
and then subtracting the corresponding backgrounds. 
In deriving these profiles, we excluded a region of radius $2'.5$ 
centered on the Seyfert 1 galaxy EXO 1346.2+2645. 
For comparison, we also show the instrumental 
PSF of the XRT+GIS combination.
Although this PSF was calculated over the total 0.7--10 keV band,
it can well approximate those in either soft or hard energy band.

In order to compare the classical soft-band brightness profiles of A1795
against the hard-band ones that have been obtained for the first time,
we have normalized the latter to the former at each radius.
The results, or radial profiles of the spectral hardness ratio,
are presented in Figure 4c for the GIS and SIS separately.
They are very flat; 
the GIS profile is almost constant within $\sim 10$\%, except beyond $\sim 8'$ 
where the instrumental artifact mentioned in \S 3.2 begins to affect the data.
The SIS hardness ratio indicates a slight drop in $r< 2'$;
this is presumably because the SIS has a higher spectral sensitivity 
and a better spatial resolution to detect the central cool emission,
whose presence was already confirmed in \S3.4. 
The constancy of the hardness-ratio profiles implies
that the central excess emission that has been observed in the soft X-ray band
should exist also in the hard X-ray band.
This inference is consistent with the results obtained in \S 3
that the spectral changes at the cluster center are rather modest.

Two obvious criticisms may be addressed to the above inference.
First, the central excess emission may have 
too soft a spectrum to be detected with \ASCA.
This is however unlikely, since the average ICM temperature of A1795
derived with \ROSAT\ through the cluster center, 2.9 keV (Briel \& Henry 1991), 
is clearly high enough for the \ASCA\ bandpass.
Alternatively, the central soft X-ray excess may be 
undetectable with \ASCA\ because of the heavy smearing by the PSF. 
In order to examine this issue in the simplest manner,
an appropriately scaled 3--10 keV radial profile of a point source
was added to, or subtracted from, the hard-band radial profile of A1795.
This is because any central excess emission
should appear almost pointlike to \ASCA.
For the point source, we used the actual GIS data of Cygnus X-1
observed at the same position as A1795 (Takahashi et al. 1995).
The hard-band profile of A1795, thus modified, 
was then divided again by the soft-band GIS profile of A1795.
The results shown in Figure 4d clearly confirm
that the hardness ratios deviate significantly from the actually measured values
when the point-source contribution (either positive or negative)
exceeds $\sim 3$\% of the A1795 counts integrated within $12'$.

>From these considerations, we conclude 
that the hardness-ratio profile obtained with \ASCA\ is very
sensitive to small differences in the fractional central brightness excess 
between the two energy bands. 
Given that the relative excess is 30--50\% in energies 
below $\sim 3$ keV (Jones \& Forman 1984; Briel \& Henry 1996),
the relative excess should be at least 25\% in hard X-rays. 

\subsection{Single-$\beta$ Fitting}

\subsubsection{The Entire Cluster Region}

To strengthen the conclusion derived in \S4.1,
we need to quantify the X-ray count-rate profiles.
However, an X-ray image obtained with \ASCA\ is a very complex 
integral transform (not a simple convolution) of the 
original surface brightness distribution on the sky plane,
because the PSF of XRT is axially non-symmetric and heavily position dependent.
Accordingly, instead of trying to inversely solve this integral transform,
we again employ the Monte-Carlo simulation used in \S 3.3.
That is, we generate simulated cluster data 
according to the $\beta$ model of equation (1),
in which the XRT+GIS PSFs are again represented by the actual GIS image of 
Cygnus X-1 observed at various detector positions (Takahashi et al. 1995).
We then derive simulated GIS count-rate profiles in several energy bands,
in the same way as has been done with the actual GIS data.
By evaluating their agreements with the actual GIS radial profiles,
we search for the best-fit $\beta$ model. 
We added a systematic error of 3\% to the simulated radial profiles.
This technique has been developed and applied to the \ASCA\ data 
by Takahashi et al. (1995), Ikebe (1995),  
Markevitch et al. (1996), Honda et al. (1996), Matsuzawa et al. (1996),
Ohashi et al. (1997), and Ikebe et al. (1997).

We first applied this $\beta$-model to the GIS radial count-rate profiles 
over the whole radius range of $r< 12'$.
As summarized in Table 2, rather flat $\beta$ models with small core radii
provided the best-fit solutions in the soft (0.7--3 keV) 
and hard (3--10 keV) energy bands. 
However, none of these fits are acceptable, 
exhibiting subtle but statistically significant fit residuals as shown in Figure 5a-5b.
These failures of the fit can be ascribed obviously 
to the presence of the central excess emission, 
as already pointed out by Jones \& Forman (1984) and Briel \& Henry (1996).

We also analyzed the processed archival \ROSAT\ PSPC data of A1795, 
provided by the \ROSAT\ Guest Observer Facility at the NASA/GSFC.  
These data, acquired on 1991 July 1 and 2,
comprise a part of the datasets used by Henry \& Briel (1996).
The pointing position is $\alpha=13^{\rm h}48^{\rm m}52.8^{\rm s}$, 
$\delta =+26^{\circ}37'12^{''}.0$ (J2000), 
just near the center of the cluster.
By using the exposure map, we corrected the broad-band (0.2--2 keV) image 
for vignetting and variations in the exposure across the field of view. 
We then derived the azimuthally-averaged radial 
count-rate profile in the 0.2--2 keV range, 
in which the Seyfert 1 galaxy (EXO 1346.2+2645) 
was omitted by a circular mask of $1.'5$ radius. 
The background is not subtracted, but considered in the latter model fitting.

We performed a single-$\beta$ model fitting to the derived \ROSAT\
count-rate profile, over the radius range of $r < 12'$.
We approximated the PSF with a delta-function,
which is justifiable when we ignore spatial structures much finer than $\sim 1'$
(H. B\"{o}hringer, a private communication).
We included a background component into the fit model
assuming it to be free but spatially constant.
As shown in Table 2 and Figure 5c, the fit turned out to be unacceptable, 
confirming the report by Briel \& Henry (1996).

In spite of these failures, we note 
that the single-$\beta$ model can approximate the observed 
X-ray profiles to a reasonable accuracy (e.g. within 5--10 \%);
this allowed us to utilize the single-$\beta$ approximation of the 
brightness profile in running the Monte-Carlo spectral simulation (\S 3.3).
It is still more important to note that the best-fit $\beta$-model parameters 
(though unacceptable) are quite similar between the \ROSAT\ PSPC and the \ASCA\ GIS,
and among the two energy bands of \ASCA\ (Table 2).
Furthermore, the fit residuals behave essentially 
in the same manner in the three panels of Figure 5:
the data exceed the model at $r=3'-6'$,
while the opposite occurs in other regions.
These results support our inference made in the previous subsection
that the count-rate profiles do not depend on the energy significantly.

\subsubsection{Outer Cluster Region}

When the central cluster region of $\sim 5'$ radius is ignored,
the single-$\beta$ model has given acceptable fits to the GIS profiles 
in the three energy bands (soft, hard, and total),
as well as to the 0.2--2 keV \ROSAT\ profile.
We present these fits in Figure 6,
and summarize the best-fit parameters in Table 3.

Clearly, the $\beta$-model parameters derived with the \ASCA\ GIS
are again nearly energy-independent within the respective errors.
They also agree with those derived with the \ROSAT\ PSPC
which in turn are close to those obtained by Briel \& Henry (1996) 
who used a larger \ROSAT\ datasets, discarding the central $\sim 5'$.
Furthermore, the previous results from the \Einstein\ IPC (Jones \& Forman 1984) 
are roughly consistent with those obtained here, as we can see in Table 3. 
In a word, regardless of the used instrument or the employed energy range,
we obtain a core radius $\sim 3'-4'$ and a slope of $\beta = 0.8-0.9$
for A1795, as long as the central $\sim 5'$ region is ignored.

When extrapolated inwards, these best-fit $\beta$ models 
fall significantly below the observed brightness (Fig. 6), 
thus reconfirming the prominent central excess emission.
The last column of Table 3 gives a rough measure of this central excess,
integrated within $r <5'$ and normalized to the 
total counts integrated up to a projected radius of $12'$.
Thus, the measurements with \Einstein, \ROSAT, and \ASCA\
consistently indicate a central excess emission
that reaches $20-30$\% of the total X-ray emission within $12'$,
although the excess may decrease mildly with energy, which is most
likely to be caused by the cool emission component.

Combining these results with those obtained in the previous subsections,
we draw the following two conclusions of basic importance.
One is that the central excess emission of A1795 it too strong
to be smeared out by the PSF of \ASCA,
and the other is that the central X-ray excess emission 
of this cluster is significantly seen 
over the entire 0.2--10 keV energy range spanned by \ROSAT\ and \ASCA. 

\subsection{Double-$\beta$ Fitting}

In order to describe the central excess emission more quantitatively,
we further attempted to fit the count-rate profiles
with a sum of two $\beta$ model components,
a more extended one representing the emission outside $\sim 5'$,
and a compact one representing the central excess.
We have found that this double-$\beta$ model gives acceptable fits 
to the profiles obtained with the \ROSAT\ PSPC and the \ASCA\ GIS,
as summarized in Table 4 and shown in Figure 7.
All the observed count-rate profiles have thus yielded very similar 
parameter values for the two $\beta$-model components.
The more extended component is similar to that obtained 
in the single-$\beta$ fits excluding the central region (Table 3),
while the compact component has a core radius 
which is close to that obtained with the single-$\beta$ fits
including the central region (Table 2).

We remark that the very central ($r < 1'$) region of the cluster 
may remain unresolved even with the \ROSAT\ PSPC. 
Consequently, the core radius of $a \sim 1'$ for the compact component 
may be subject to instrumental angular resolutions,
and the finer-resolution X-ray brightness profiles,
e.g. those obtained with the \ROSAT\ HRI,
may no longer be adequately described by our double-$\beta$ model.
Nevertheless, our principal conclusions should remain unchanged,
because they all deal with angular scales of $\sim 1'$ or larger.

In Table 5, we summarize the luminosity, $L_{\rm excess}$, 
carried by the narrower $\beta$ component. 
It amounts to $40-50$\% of the total (i.e., narrow plus wide) 
luminosity, $L_{\rm total}$, in both the soft and hard energy bands.
This is reflected in the fact that the two $\beta$ components crossing over in
Figure 7 have similar normalization ratios in the two \ASCA\ energy bands.
The excess luminosity defined in this way is systematically larger 
than those given in Table 3, by up to a factor of $\sim2$. 
However, this is simply the matter of how to define the ``excess''.

\subsection{Fitting with the Universal Halo Model}

Through extensive $N$-body numerical experiments of structure formation 
in the cold dark matter (CDM) scenario, Navarro, Frenk, \& White (1996) reported 
that the computed CDM mass density $\rho_{\rm DM}$ takes a ``universal halo shape'' of 
\begin{equation} 
    \rho_{\rm DM}(R) 
       = \frac{\delta}{ \left( \frac{R}{R_{\rm s}} \right) 
                         \left( 1+\frac{R}{R_{\rm s}} \right)^2}
\end{equation}
where $\delta$ and $R_{\rm s}$ are parameters.
Makino, Sasaki, \& Suto (1997) showed 
that the density of an isothermal gas hydrostatically confined 
in the potential of equation (2) can be expressed analytically as 
\begin{displaymath} 
     n(R) = n_0 \ \left(1+ \frac{R}{R_{\rm s}}\right)^{\frac{BR_{\rm s}}{R}}
\end{displaymath}
where $n_0$ and $B$ are also parameters.
By numerically projecting the square of this $n(R)$ onto 2 dimensions,
we can obtain the X-ray surface brightness, $S_{\rm UH}(r)$,
emitted by an isothermal gas sphere confined in a universal CDM halo.

Since the CDM distribution of equation (2) is more centrally peaked 
than the King-type distribution, 
the idea of universal CDM halo may explain the central excess of A1795.
In order to examine this possibility,
we fitted this $S_{\rm UH}(r)$ to the \ROSAT\ PSPC profile.
However as shown in Figure 7d, the fit turned out to be as poor as 
that with the single-$\beta$ model (Fig.5c), with $\chi^2/\nu= 56/16$.
The GIS profiles could not be fitted with this new model, either.
This result is not surprising,
since $S_{\rm UH}(r)$ is in fact fairly close to the 
ordinary $\beta$-model profile $S(r)$ given by equation (1),
with only moderate excess towards the center (Makino, Sasaki, \& Suto 1997). 
In short, the central excess emission from A1795 is too strong
to be explained by the universal halo model as long as the isothermality holds.

\section{DISCUSSION}

\subsection{Summary of the Results}

Through a series of analysis of the radial count-rate profiles 
of A1795 obtained with the \ASCA\ GIS and the \ROSAT\ PSPC,
we have confirmed that the central excess brightness of this cluster, 
which was well known previously in the soft X-ray band,
does exist also in the 3--10 keV band with a similar fraction.
If employing $\beta$-models, two distinct components are required 
to express the projected radial count-rate profiles obtained 
with the \ASCA\ GIS in the energy range both below and above 3 keV, 
as well as that obtained with the \ROSAT\ PSPC.

We also carried out spectroscopic studies 
of the ICM in A1795 using the \ASCA\ data. 
In essence, the ICM in A1795 is isothermal at a temperature of 6.2 keV,
except in the innermost region of radius $\sim 3'$ ($\sim 320$ $h_{50}^{-1}$ kpc) 
where an additional cool emission component is seen
as already report by Fabian et al. (1994).
Employing the two-temperature fit obtained in \S 3.4,  
we calculated the luminosities of the cool component 
in various energy bands, as listed in Table 5.

In the 0.5--3 keV band, these morphological and spectroscopic results compare 
as $L_{\rm cool}/L_{\rm excess}=0.42 \pm 0.16$ (according to Table 3),
or $0.24 \pm 0.07$ (according to Table 5).
These ratios may not necessarily imply a discrepancy 
between $L_{\rm cool}$ and $L_{\rm excess}$, 
since our estimation of $L_{\rm cool}$ depends on
a particular modeling (i.e., two-temperature assumption) of the ICM.
However the discrepancy becomes much more significant 
when the comparison is made in the newly explored 3--10 keV band;
$L_{\rm cool}/L_{\rm excess}=0.15\pm 0.04$ in reference to Table 3, 
or $0.07 \pm 0.02$ in reference to Table 5.
These small ratios directly result from the fact
that the spectral cool component has a temperature of at most $\sim 2$ keV (Table 1),
whereas the morphological excess component has a temperature 
rather close to the cluster-average value of 6.2 keV.

Our results consistently indicate 
that the whole cluster volume of A1795 is permeated with the hot isothermal ICM, 
including the central region where a small amount of cool component coexists.
This is similar to the case of several other cD clusters
(Fukazawa et al. 1994; Ikebe 1995;  Matsumoto et al. 1996; Ikebe et al. 1997).
The central excess brightness of A1795 is inferred to be produced primarily 
by a central excess in the density, hence pressure, of this hot ICM itself,
with the cool emission component giving subsidiary effects.
To confine this excess ICM pressure, the gravitational potential in A1795 
must exhibit an additional central drop as compared to the King-type potential.
As indicated by Figure 7d, this central drop may be much 
deeper than is predicted by the universal halo model of equation (2),
as long as the isothermality approximation is valid.

\subsection{Integrated Mass Profiles}
Referring to the double-$\beta$ solutions 
which provide acceptable fits to the observed brightness profiles,
we quantify the shape of the gravitational potential.
Specifically, we employ the best-fit double-$\beta$ model 
obtained in the GIS 0.7--10 keV band (Table 4),
and back-project the constituent $\beta$-model 
components individually into 3-dimensions.
This provides emissivity profiles of the extended and compact components,
denoted as $\epsilon_{\rm e}(R)$ and  $\epsilon_{\rm c}(R)$, respectively. 
These can be converted to the ICM density profile $n(R)$ through the relation
\begin{equation} 
 \epsilon_{\rm e}(R) + \epsilon_{\rm c}(R) = \Lambda(T, A) n(R)^2 ~~,
\end{equation}
where $R$ is the 3 dimensional radius and $\Lambda(T, A)$ is the 
cooling function for a given temperature $T$ and abundance $A$.
We then calculate the radially integrated total gravitating mass, 
$M_{\rm tot}(R)$, using the usual equation of
\begin{equation} 
M_{\rm tot}(R) = -\frac{k_{\rm B} T R}{\mu m_{\rm p} G}
 \left\{  \frac{{\rm dln}n(R)}{{\rm dln}{R}}
       + \frac{{\rm dln}T(R)}{{\rm dln}{R}} \right\}~.
\end{equation}

The total mass profiles calculated via equation (4) are presented in Figure 8, 
where we assumed two alternative cases for the temperature profiles $T(R)$;
either a constant ICM temperature at 6.2 keV, 
or a centrally decreasing profile of the form of 
\begin{equation} 
   T(R) =  4.50  + 0.34 R -1.57 \exp \left[ -0.67 R \right] ,  
\end{equation}
which approximately reproduces the \ROSAT\ PSPC result of Briel and Henry (1996).
Here the 3 dimensional radius $R$ is in arcminute 
and the temperature $T(R)$ is in keV.
We also present the ICM mass profile, $M_{\rm ICM} (R)$, 
calculated by integrating $n(R)$ of equation (3).

Inside $R \sim 50-100 ~ h_{50}^{-1}$ kpc,
the mass curves in Figure 8 should be regarded as 
subject to the limited angular resolution. 
Finer details of the central density profile
could be studied using the \ROSAT\ HRI data, 
but the lack of associated spectral information
hampers the calculation of gravitational potential there. 
In this paper, we therefore limit our study to the information 
available with the \ROSAT\ PSPC and \ASCA.

Figure 8 reveals a shoulder-like structure 
in the $M_{\rm tot}(R)$ profile at $R_{\rm X}\sim 150 h_{50}^{-1}$ kpc,
regardless of the assumed temperature profile.
As a result,
the total gravitating mass within $R_{\rm X}$ apparently exhibits 
an excess by $M_{\rm excess} \simeq 3 \times 10^{13}~ M_{\odot}$,
above the large-scale smooth mass distribution 
carried by the extended $\beta$-component.
This $M_{\rm excess}$ is thought to be responsible 
for the central dimple in the gravitational potential.

\subsection{Nature of the Central Excess}

The shoulder-like structure seen in the $M_{\rm total}$ curve of 
Figure 8 may allow two slightly different interpretations.
One possibility is that the overall mass distribution is made up 
of a single mass component, presumably dominated by dark matter,
whose density keeps rising to the center instead of forming a flat core.
Although equation (2), representing one particular example of such cases, 
failed to explain the X-ray data under the isothermal assumption, 
some other numerical simulations (Moor et al. 1997) suggest
even a steeper CDM halo profile.
When taking into account the effect of the cool component
and baryonic mass associated with the cD galaxy,
such a centrally-steepening CDM halo may still 
be consistent with the observation.

Alternatively, the $M_{\rm tot}(R)$ profile in Figure 8 may be interpreted as 
exhibiting a gravitational hierarchy between the overall cluster and the cD galaxy.
Such a nested potential structure was suggested 
or assumed previously (e.g. Thomas et al. 1987),  
and has been confirmed in the Fornax cluster (Makishima 1995; Ikebe et al. 1996),
the elliptical galaxy NGC~4636 (Matsushita 1997; Matsushita et al. 1996, 1997),
and possibly in about 8 galaxy groups (Mulchaey \& Zabludoff 1997).
Then, the region inside $R_{\rm X}$ is considered to belong to the cD galaxy,
while the outside region to the entire cluster.
In reference to the above $M_{\rm excess}$ and the optical B-band luminosity
of the cD galaxy ($6.5\times10^{11}$ in unit of the solar B--band luminosity), 
the mass-to-light ratio within $R_{\rm X}$ becomes $\sim 50$.

The hierarchical potential structure is also suggested 
by the optical observations by Johnstone, Naylor, \& Fabian (1991), 
who reported that the {\it I}-band surface brightness of A1795 consists 
of two components (though the extended one is quite elongated),
crossing over at $R_{\rm opt}~^{>}_{\sim}100 h^{-1}_{50}$ kpc;
this is close to $R_{\rm opt}$.
Furthermore, $M_{\rm excess}$ estimated above
is similar to the virial mass of the cD galaxy 
($M_{\rm virial} \simeq 1.1\times10^{13}~ M_{\odot}$),
derived from the stellar velocity dispersion ($\simeq 300$ km sec$^{-1}$; 
Oegerle \& Hoessel 1991; McNamara \& O'Connell 1993). 
In comparison, we have  
$R_{\rm X} \sim R_{\rm opt} \sim 60~ h^{-1}_{50}$ kpc and 
$M_{\rm excess} \sim M_{\rm virial }\sim 10^{12}~ M_{\odot}$ in the Fornax cluster.

The hierarchy interpretation relies much on the fact
that the logarithmic curvature of $M_{\rm tot}(R)$ becomes apparently concave 
over the region where the shoulder-like structure is present ($R = 150-350$ kpc).
However, this may be a model dependent artifact, produced e.g. by cross-over 
between the two components in the employed double-$\beta$ model.
As a matter of fact, it is rather difficult to assess the reality of this feature,
due to complex error propagation from the data to the mass curve.
Instead of attempting to do so, we here resort to fit the observed 0.7--10 keV GIS
brightness profile by a rather arbitrary empirical function of the form
\begin{equation} 
   S(r) = S(0) \exp\left[-\left( \frac{r}{r_{1}} \right)^{c_{1}} \right] 
            \left[ 1+ \left( \frac{r}{r_{2}} \right)^2 \right]^{-c_{2}}   
\end{equation}
where $S(0)$, $r_{1}$, $r_{2}$, $c_{1}$, and $c_{2}$ are parameters.
The best-fit of this empirical model is found with $r_{1} = 7'.72$,
$r_{2} = 0'.60$, $c_{1} = 1.67$, and $c_{2} = 1.05$, yielding $\chi^{2}/\nu = 23/15$,
which falls between those from the single-$\beta$ fit and the double-$\beta$ fit. 
The mass profile derived from this ad-hoc model,
via back projection and equation (4), is also plotted in Figure 8. 
It reveals a very similar feature with concave curvature,
even though equation (6) does not involve 
particular angular scales corresponding to $2'-3'$
and the logarithmic derivative of equation (6) changes monotonically with $r$. 
This adds to the reliability of the concave feature seen in Figure 8,
and reinforce the interpretation of our results in terms of the
hierarchical potential structure between the cluster and the cD galaxy.

Finally, some mention may be made on cooling flows.
The dominance of hot ICM in the cluster center region, 
with a smaller amount of cooler component coexisting with it,
is in itself consistent with the picture of 
inhomogeneous cooling flows (Fabian 1994). 
The short cooling time of 2.5 Gyr within the central 80 kpc,
as estimated by Edge, Stewart, \& Fabian (1992),
remains essentially unchanged by our results.
However, causal relation between the nested potential structure revealed here
and the cooling flows in A1795 remains unclear. 
The hierarchical mass distribution may have formed first 
during the cluster formation, 
and provided a favorable environment for cooling flows to take place.
If so, the cooling flow may be regarded as the result, rather than the cause,
of the central excess X-ray emission.
Alternatively, a certain fraction of the excess mass within $R_{\rm X}$
may have deposited from the cooling flow itself.

In summary, we have discovered that the central excess X-ray emission 
from A1795 is seen in the hard X-ray band above 3 keV, 
with a similar fraction as seen in the soft X-rays.  
This property is mainly due to a central excess in the gravitating mass, 
by $M_{\rm excess} \sim 3 \times 10^{13}~ M_\odot$,
which is likely to be associated wit the cD galaxy.

\vspace{1cm}
We thank all the members of the \ASCA\ team.
We would also like to thank Hans B\"{o}hringer, Wolfgang Voges, 
and Doris Neumann for their helpful discussions and comments 
on the analysis of the \ROSAT\ PSPC data. 
We are also grateful to A. C. Fabian, R. F. Mushotzky,
S. Sasaki, and Y. Suto for helpful comments.
This research has made use of data obtained through the High Energy
Astrophysics Science Archive Research Center Online Service, 
provided by the NASA-Goddard Space Flight Center.


\newpage

\section*{References}

\begin{list}{}{\setlength{\leftmargin}{3em}\setlength{\rightmargin}{0cm}
\setlength{\itemindent}{-2em}
\setlength{\itemsep}{3pt}
\setlength{\baselineskip}{15pt}}

\item Anders, E., \& Grevesse, N. 1989, Geochim. Cosmochim. Acta 53, 197 
\item Briel, U. G., \& Henry, J. P. 1996, ApJ, 472, 131
\item Burke, B. E., Mountain, R. W., Daniels, P. J., \& Dolat, V. S. 1994,
      IEEE Trans. Nuc. Sci., 41, 375
\item David, L. P., Arnaud, K. A., Forman, W., \& Jones, C. 1990, ApJ, 356, 32
\item Edge, A. C., \& Stewart, G. C. 1991a, MNRAS, 252, 414
\item Edge, A. C., \& Stewart, G. C. 1991b, MNRAS, 252, 428
\item Edge, A. C., Stewart, G. C., \& Fabian, A. C. 1992, MNRAS, 258, 177
\item Fabian, A. C. 1994, ARA\&A, 32, 277
\item Fabian, A. C., Arnaud, K. A., Bautz, M. W., \& Tawara, Y. 1994, 
      ApJ, 436, L63
\item Fukazawa, Y., Ohashi, T., Fabian, A. C., Canizares, C. R., Ikebe, Y., 
      Makishima, K., Mushotzky, R. F., \& Yamashita, K. 1994, PASJ, 46, L55
\item Hatsukade, I. 1989, Ph.D. thesis, Osaka University
\item Honda, H., et al. 1996, ApJ, 473, L71
\item Ikebe, Y. 1995, Ph.D. thesis, Tokyo University
\item Ikebe, Y., et al. 1996, Nature, 379, 427
\item Ikebe, Y., et al. 1997, ApJ, 481, 660 
\item Johnstone, R. M., Naylor, T., \& Fabian, A. C. 1991, MNRAS, 248, p18
\item Jones, C., \& Forman, W. 1984, ApJ, 276, 38
\item King, I. 1972, ApJ, 174, L123
\item Makino, N., Sasaki, S., \& Suto, Y. 1997, ApJ., in press
\item Makishima, K. 1995, in Dark Matter (AIP Proceedings No.336),
      ed. S. S. Holt, \& C. Bennet (New York: AIP), p.172
\item Makishima, K. 1996, in UV and X-Ray Spectroscopy of Astrophysical 
      and Laboratory Plasmas, ed. K. Yamashita, \& T. Watanabe  
      (Tokyo: Universal Academy Press), 167
\item Makishima, K., et al. 1996, PASJ, 48, 171
\item Mulchaey, J. S., \& Zabludoff, A. I. 1997, ApJ, in press
\item Markevitch, M., Mushotzky, R., Inoue, H., Yamashita, K., Furuzawa, A., \&
      Tawara, Y. 1996, ApJ, 456, 437
\item Matsumoto, H., Koyama, K., Awaki, H., Tomida, H., Tsuru, T., Mushotzky, R., 
                                         \& Hatsukade, I. 1996, PASJ, 48, 201
\item Matsushita, K. 1997, PhD Thesis, University of Tokyo
\item Matsushita, K., et al.  1996, in The Nature of Elliptical Galaxies 
      (Proceedings of the Second Stromlo Symposium),
      ed. M. Arnaboldi, G. S. Da Costa, \& P. Saha, 407
\item Matsushita, K., et al.  1997, submitted to Nature
\item Matsuzawa, H., et al. 1996, PASJ, 48, 565
\item McNamara, B. R., \& O'Connell, R. W. 1993, A\&A, 270, 60
\item Miralda-Escude, J., \& Babul, A. 1995, ApJ, 449, 18
\item Moore, B., Governato, F., Quinn, T., Stadel, J., \& Lake, G. 1997, submitted to ApJL
\item Mushotzky, R., Loewenstein, M., Arnaud, K., \& Fukazawa, Y. 1995,
      in Dark Matter (AIP Proceedings No.336),
      ed. S. S. Holt, \& C. Bennet (New York: AIP), 231
\item Mushotzky, R. 1994, in New Horizon of X-ray Astronomy,
      ed. F. Makino, \& T. Ohashi (Tokyo: Universal Academy Press), 243
\item Navarro, J. F, Frenk, C. S, \& White, S. D. M. 1996, ApJ, 462, 563
\item Oegerle, W. R., \& Hoessel, J. G. 1991, ApJ, 375, 15
\item Ohashi, T, H. Honda, H. Ezawa, \& K. Kikuchi 1997, 
      in X-ray Imaging Spectroscopy of Cosmic Hot Plasmas,
      ed. F. Makino, \& K. Mitsuda (Tokyo: Universal Academy Press), 49 
\item Ohashi, T., et al. 1996, PASJ, 48, 157
\item Raymond, J. C., \& Smith, B. W. 1977, ApJS, 35, 419 
\item Sarazin, C. L. 1988, X-Ray Emission from Clusters of Galaxies 
                                                (Cambridge: Cambridge University Press) 
\item Serlemitsos, P. J., et al. 1995, PASJ, 47, 105 
\item Schombert, J. M. 1986, ApJS, 60, 603
\item Takahashi, T., et al. 1995, \ASCA\ Newsletter (NASA/GSFC), 3, 25 
\item Tanaka, Y., Inoue, H., \& Holt, S.S. 1994, PASJ, 46, L37
\item Thomas, P., Fabian, A. C., \& Nulsen, P. E. J. 1987, MNRAS, 228, 973
\item Tsuru, T. 1992, Ph.D. thesis, Tokyo University
\item Tsusaka, Y., et al. 1995, Appl. Opt., 34, 4848   
\item White, D. A., Fabian, A. C., Johnstone, R. M., Mushotzky, R. F., 
      \& Arnaud K. A. 1991, MNRAS, 252, 72 
\item Wu, X., \& Hammer, F. 1993, MNRAS, 262, 187

\end{list}

\newpage
\noindent
{\bf\LARGE Figure captions}

\vspace{7mm}

\noindent
Figure 1:  
Background-subtracted GIS2 + GIS3 images of A1795 
in (a) 0.7--3 keV and (b) 3--10 keV.
The images have been smoothed with two-dimensional Gaussian filters ($\sigma=0'.5$). 
No correction has been made for the XRT vignetting 
or partial shadows due to the detector support ribs. 
The contours are with the same logarithmic steps in the two images. 
The SIS field of view is indicated by the dashed line in each plot, 
and the position of the Seyfert 1 galaxy EXO 1346.2+2645 is marked by a cross.
           
\vspace{5mm}

\noindent
Figure 2: 
(a) Spatially sorted GIS2 + GIS3 spectra of A1795 for $0'-2'$ and $2'-12'$ regions. 
The background has been subtracted, but the detector response has not been removed. 
(b) Same as (a) but for the SIS0 + SIS1 spectra.

\vspace{5mm}

\noindent
Figure 3: 
(a) The GIS spectrum in the $r<2'$ region (same as shown in Fig.2a),
displayed in the form of channel-by-channel ratios to the simulated GIS spectrum.
The Monte-Carlo simulation was done taking fully into account the PSF effects,
assuming a constant ICM temperature of 6.2 keV
and a constant abundance of 0.3 solar. 
The model birghtness profile was represented by a single $\beta$-model 
with a core radius of $0.'8$ and  $\beta= 0.56$.
Eerror bars represent $\pm 1 \ \sigma$ due to 
photon counting statistics in the actual spectrum, 
while those associated with the simulated spectrum is negligbly small. 
(b) Same as (a), but for the $r= 6'-8'$ region.

\vspace{5mm}

\noindent
Figure 4:
(a) Background-subtracted radial count-rate profiles of A1795 
obtained with GIS2 + GIS3 in 0.7--3 keV and 3--10 keV.
The profiles are corrected for neither the XRT vignetting nor the XRT PSF effects. 
The 0.7--10 keV XRT PSF at the detector center is also plotted for comparison. 
(b) Same as (a) but for SIS0.
(c) The radial profiles of the 3--10 keV to 0.7--3 keV brightness ratio 
(or the spectral hardness ratio) of A1795, derived with GIS2 + GIS3 and SIS0. 
(d) The same GIS ratio as in (c), 
but with a 3--10 keV radial profile of a point source, 
whose integrated count is 3\% of that of total cluster, 
added to or subtracted from the 3--10 keV GIS data.  

\vspace{5mm}

\noindent
Figure 5:
Single-$\beta$ fittings to the radial count-rate profiles of A1795 
over the whole radius range of $r< 12'$.
(a) The \ASCA\ GIS profile in 0.7--3 keV, 
where crosses represent the data
while circles show the best-fit model convolved with the XRT + GIS PSF.
The lower panel shows the fit residuals.
(b) The same as panel (a) but, in the 3--10 keV range.         
(c) The \ROSAT PSPC profile in 0.2--2 keV fitted in the same manner.

\vspace{5mm}

\noindent
Figure 6:
The same as Figure 5, but the fit range is restricted to 
the outer region of $r^{>}_{\sim} 5'$.
          
\vspace{5mm}

\noindent
Figure 7:
Two-$\beta$ fittings to the brightness profiles of A1795 over the whole radius range of $r< 12'$.
Meanings of panels (a) through (c) are the same as those in Figure 5.
Squares and triangles show the wider and narrower $\beta$-components, 
respectively, while circles indicate their sum.
Panel (d) is the fit to the \ROSAT\ PSPC data,
using the X-ray surface brightness profile
which is predicted by the universal halo model 
of equation (2) in the text.

\vspace{5mm}

\noindent
Figure 8:
Radial distributions of the integrated total gravitating mass and the ICM mass. 
Cases 1 and 2 are based on the best-fit double-$\beta$ model 
in 0.7--10 keV (Table 4),
assuming either a constant ICM temperature of 6.2 keV (case 1)
or a temperature gradient given by equation (5) in the text (case 2).
Case 3 is derived by replacing the double-$\beta$ model 
with another analytical surface-brightness model 
described with equation (6) in the text,  
with the temperature assumed to be constant at 6.2 keV.

\newpage

\begin{table} 
  \caption{Single-temperature (1T) and two-temperature (2T) Raymond-Smith fits 
           to the SIS and GIS spectra for the innermost region of $3'$ radius.$^{*}$}
  \begin{center}
    \begin{tabular}{lccccc} 
\hline 
\hline
   Model & $N_{\rm H}$& $kT_{\rm hot}$&$kT_{\rm cool}$&Abundance&$\chi^2/\nu$\\
         & ($10^{20}$cm$^{-2}$) &  (keV)&(keV)        & (solar) \\
\hline 
1T &3$^{\dagger}$ &    6.2 fixed      & --- &  0.3$^{\dagger}$ & 645/278 \\
1T & $3\pm 1$ &$5.16 \pm 0.11$        & --- & $0.34 \pm 0.02$   & 453/277 \\
2T $^{\ddagger}$
   & $5\pm 1$ &$6.54^{+0.56}_{-0.50}$ &$1.70^{+0.24}_{-0.26}$   & $0.32 \pm 0.03$ &
377/275\\
\hline
    \end{tabular}
 \end{center}
 \begin{description}
 \begin{footnotesize}
 \setlength{\itemsep}{-1mm}
  \setlength{\itemindent}{-3mm}
\item[$*$] The SIS and GIS spectra are fitted simultaneously.
\item[$^\dagger$] Errors were not calculated because of too poor a fit.
\item[$^\ddagger$] 
        Two-temperature (2T) fits, where the two components are   
                                                  assumed to have the same abundances and a common absorption. 
        Abundance rataios are assumed to be solar.
        Errors represent the single-parameter $90$\% confidence limits.       
 \end{footnotesize}
 \end{description}
\end{table}

\begin{table} 
  \caption{Single-$\beta$ fits to the count-rate profiles
           in the entire $r^{<}_{\sim}12'$ range.$^*$}
  \begin{center}
    \begin{tabular}{ccccc} 
\hline
\hline
      Detector   & Energy (keV)& Core radius & $ \beta$ & $\chi^2/\nu$ \\  
\hline
 \ASCA\ GIS     & $0.7-10$    & $0'.9 $    & $0.62 $ & 45.5/17 \\
                & $0.7-3$     & $0'.8 $    & $0.60 $ & 45.9/17 \\
                & $3-10$      & $1'.0 $    & $0.61 $ & 29.6/17 \\ 
 \ROSAT\ PSPC   & $0.2-2$     & $0'.8 $    & $0.56 $ & 49.4/16 \\
\hline
\end{tabular}
 \end{center}
  \begin{description}
  \begin{footnotesize}
  \setlength{\itemsep}{-1mm}
  \setlength{\itemindent}{-3mm}
  \item[$*$] Confidence ranges of the parameters are not shown,
             since none of the fits are acceptable.     
  \end{footnotesize}
  \end{description}
\end{table}

\begin{table}  
  \caption{Single-$\beta$ fits to the radial count-rate profiles
           in the outer region ($r^{>}_{\sim}5'$), with conservative errors.}
  \begin{center}
    \begin{tabular}{cccccc} 
\hline 
\hline 
   Detector & Energy (keV) &core radius  & $\beta$  &  $\chi^2/\nu$  & central excess$^{*}$\\
\hline
\ASCA\ GIS      &$0.7-10$   & $3'.1\pm0'.1$ & $0.81 \pm 0.05$ & 6.4/8 & 0.22 \\
                &$0.7-3$    & $3'.2\pm0'.1$ & $0.79 \pm 0.05$ & 6.2/8 & 0.28 \\
                &$3-10$     & $3'.1\pm0'.1$ & $0.80 \pm 0.05$ & 2.5/8 & 0.20 \\ 
\ROSAT\ PSPC    &$0.2-2$    & $3'.8\pm0'.3$ & $0.90 \pm 0.03$ & 4.3/9 & 0.32 \\
\Einstein\ IPC$^\dagger$  
                &$0.5-3$  & $2.'79\pm0.'93$  &  $0.72 \pm 0.08$  & ---    & 0.36 \\
\hline 
    \end{tabular}
  \end{center}
  \begin{description}
  \begin{footnotesize}
  \setlength{\itemsep}{-1mm}
  \setlength{\itemindent}{-3mm}
     \item[$^*$] Data excess in $r^{<}_{\sim}5'$ above the best-fit model,
                 relative to the total data counts in $r<12'$. 
     \item[$^\dagger$] From Jones \& Forman 1984.       
  \end{footnotesize}
  \end{description}
\end{table}

\begin{table} 
  \caption{Double-$\beta$ fits to the $r<12'$ brightness profile,
           with conservative errors.}
  \begin{center}
    \begin{tabular}{ccccccc} 
\hline
\hline
              &             & \multicolumn{2}{c}{Compact component}
                            & \multicolumn{2}{c}{Extended component}& \\
    Detector  &Energy (keV) & Core radius    & $\beta$
                            & Core radius    & $\beta$         & $\chi^2/\nu$\\   
\hline
 \ASCA\ GIS  &  $0.7-10$    & $0'.9 \pm 0'.1$ & $0.83 \pm 0.05$
                            & $3'.2 \pm 0'.1$ & $0.81 \pm 0.05$ & 10/14 \\
              &  $0.7-3$    & $1'.0 \pm 0'.1$ & $0.90 \pm 0.05$
                            & $3'.3 \pm 0'.1$ & $0.80 \pm 0.05$ & 10/14 \\
              &  $3-10$     & $0'.8 \pm 0'.1$ & $0.79 \pm 0.05$
                            & $3'.4 \pm 0'.1$ & $0.81 \pm 0.05$ & 8/14 \\ 
\ROSAT\ PSPC & $0.2-2$      & $1'.0 \pm 0'.1$ & $0.83 \pm 0.01$
                            & $3'.4 \pm 0'.1$ & $0.80 \pm 0.01$ & 7/13 \\
\hline
    \end{tabular}
  \end{center}
\end{table}

\begin{table} 
  \caption{Luminosities of the central excess, the cool component, and the total
cluster emission.$^{*}$}
  \begin{center}
    \begin{tabular}{ccccc} 
\hline 
\hline 
   Energy (keV) & $L_{\rm total}^\dagger$ &   $L_{\rm excess} ^\ddagger $ 
                  & $L_{\rm excess}/L_{\rm total}$ & $L_{\rm cool} $ \\ \hline
$0.5-10$   & $17.80\pm0.12$ & $8.37\pm0.54$ & $0.47\pm0.03$ &  $1.43 \pm 0.39$  \\
$0.5-3$    & $10.00\pm0.07$ & $4.92\pm0.30$  & $0.49\pm0.03$ & $1.20 \pm 0.33$  \\
$3-10$     & $7.80\pm0.05$  & $3.45\pm0.31$  & $0.44\pm0.04$ & $0.23 \pm 0.06 $  \\ 
\hline 
    \end{tabular}
  \end{center}
  \begin{description}
  \begin{footnotesize}
  \setlength{\itemsep}{-1mm}
   \setlength{\itemindent}{-3mm}
     \item[$^*$] Luminosities are derived through the combined fitting of
                       the \ASCA\ GIS and SIS data, in unit of
                       $10^{44}~h^{-2}_{50}$ erg s$^{-1}$ with 90\% confidence errors.
     \item[$^\dagger$] Integrated within $12'$ of the centroid.
     \item[$^\ddagger$] $L_{\rm excess}$ refers to the luminosity of the narrower
                       component in Table 4.
  \end{footnotesize}
  \end{description}
\end{table}

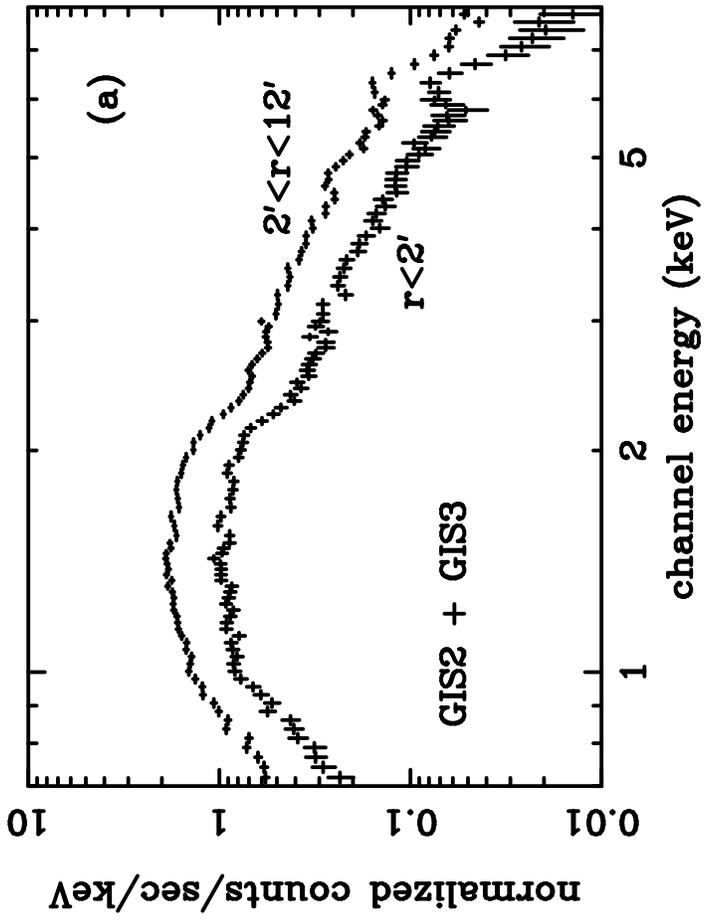
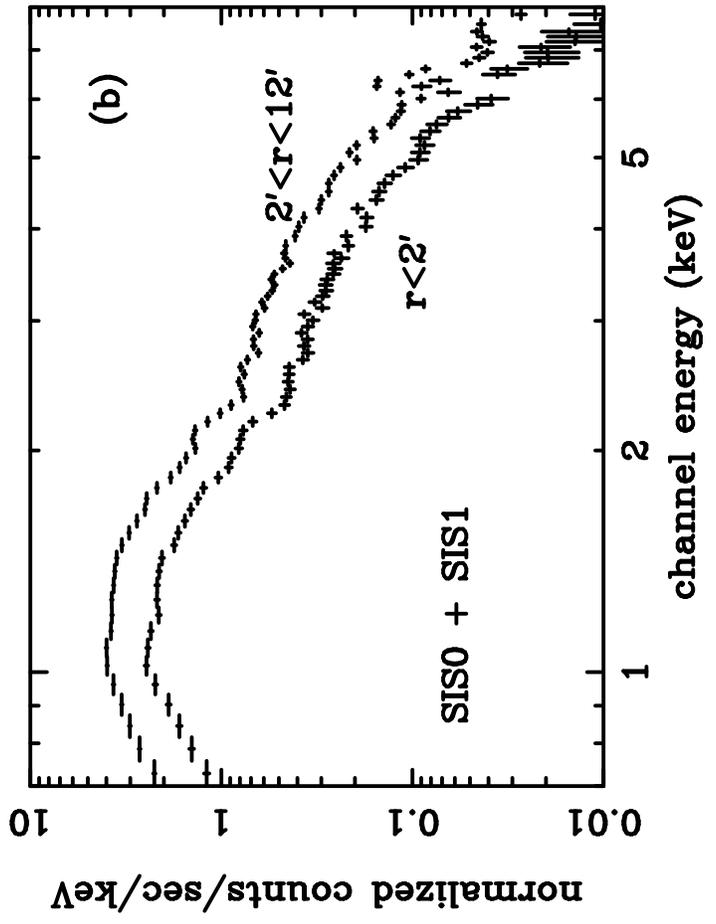
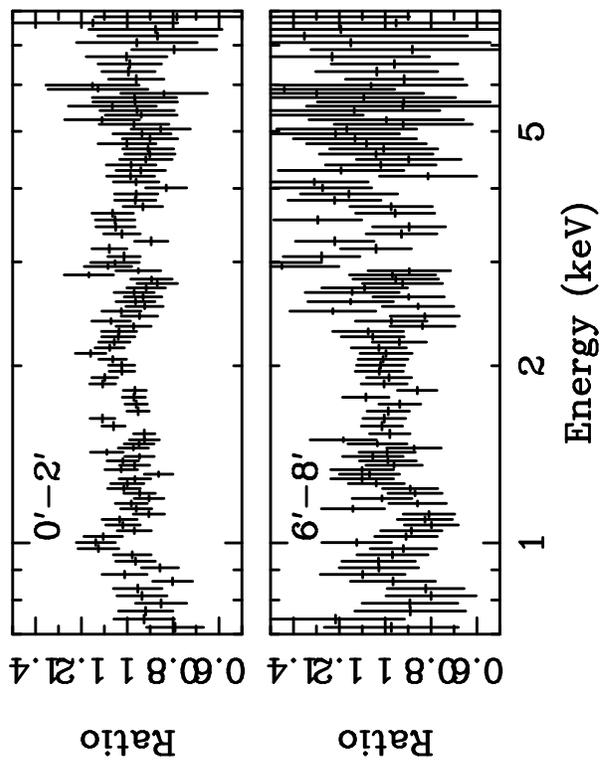
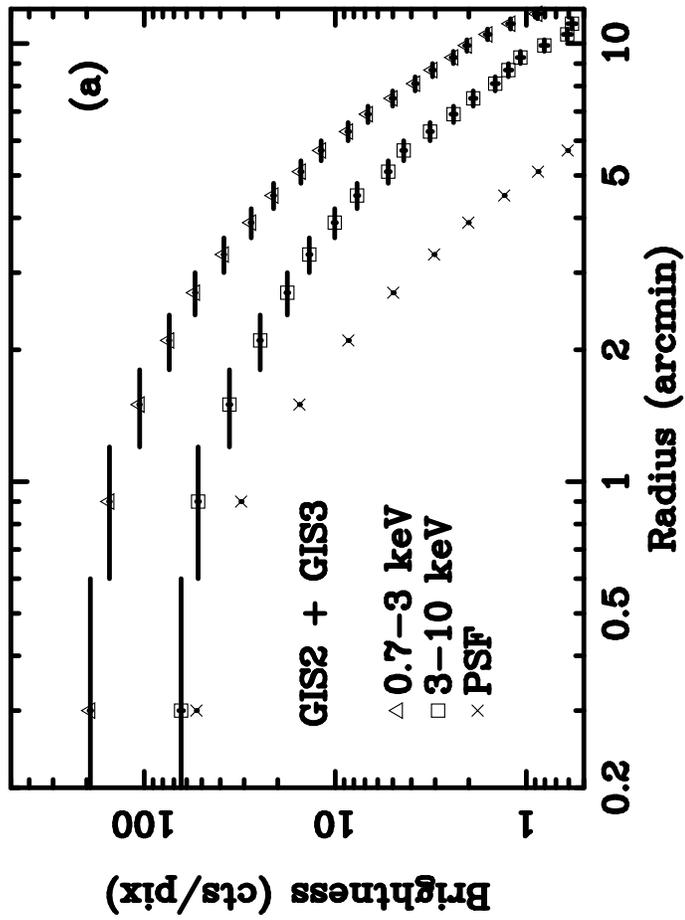
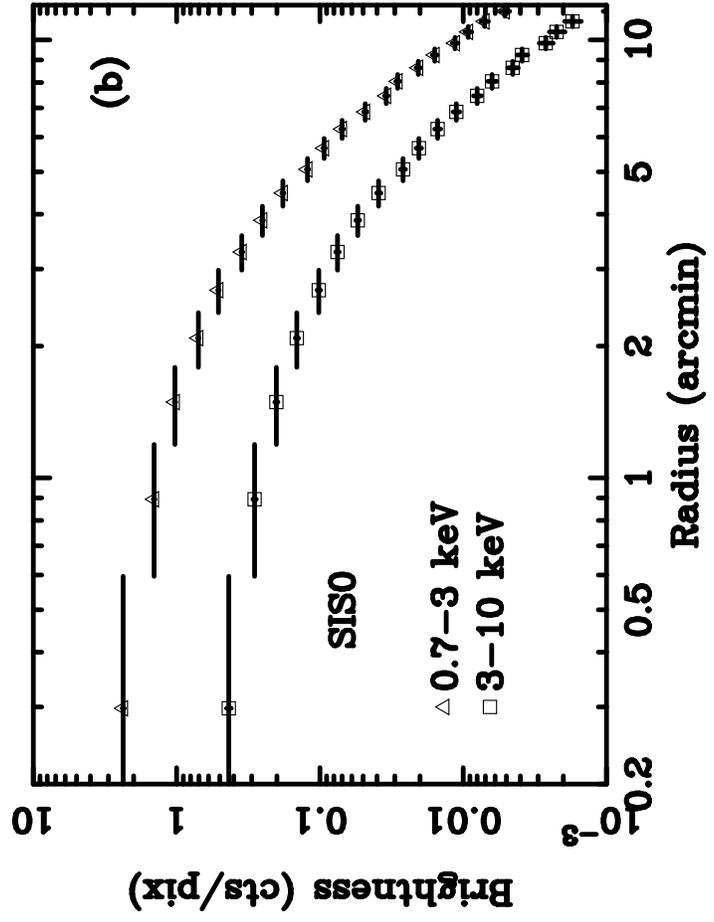
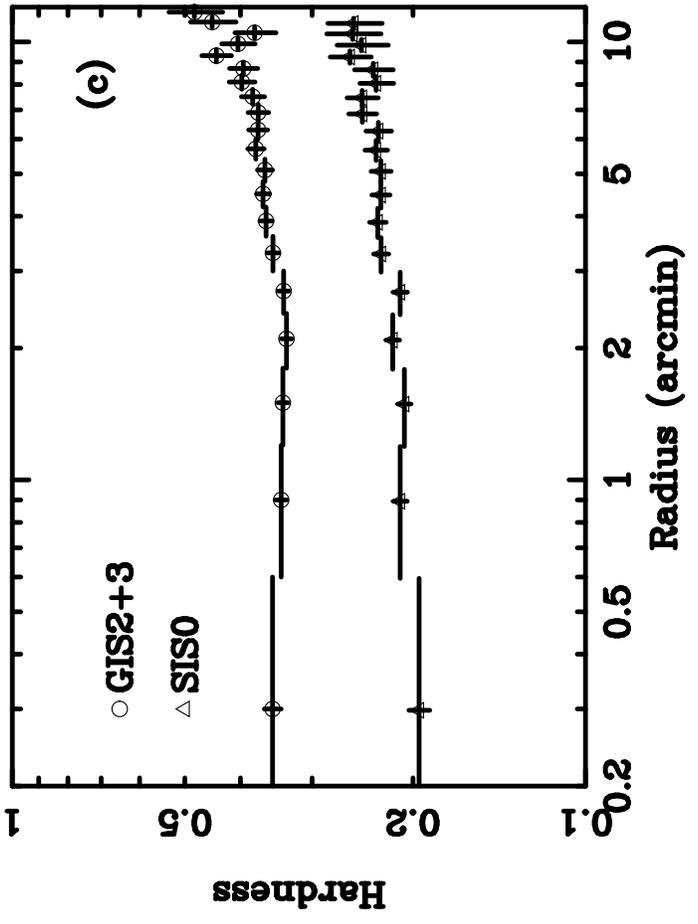
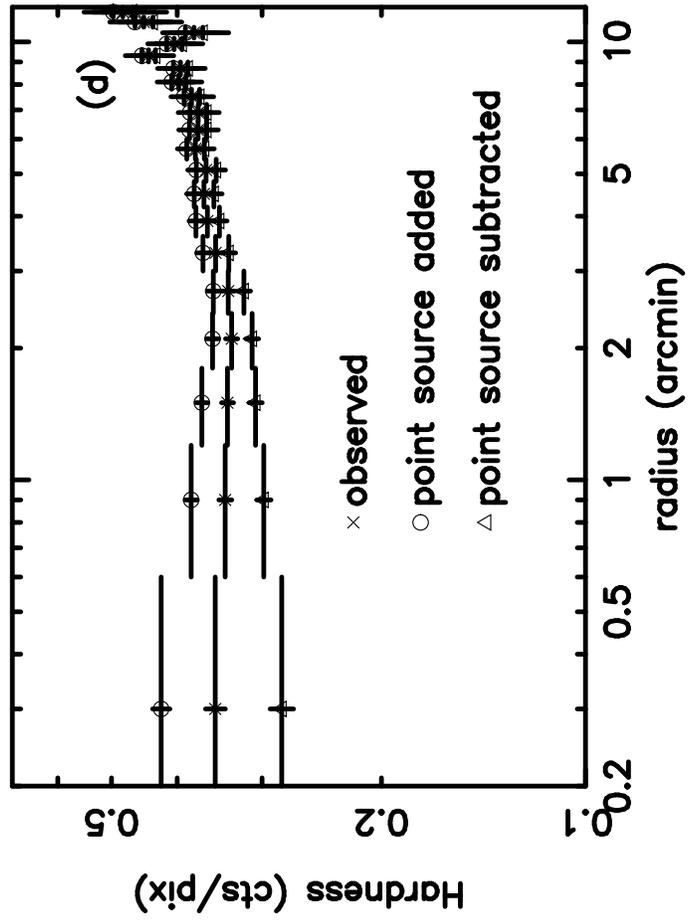
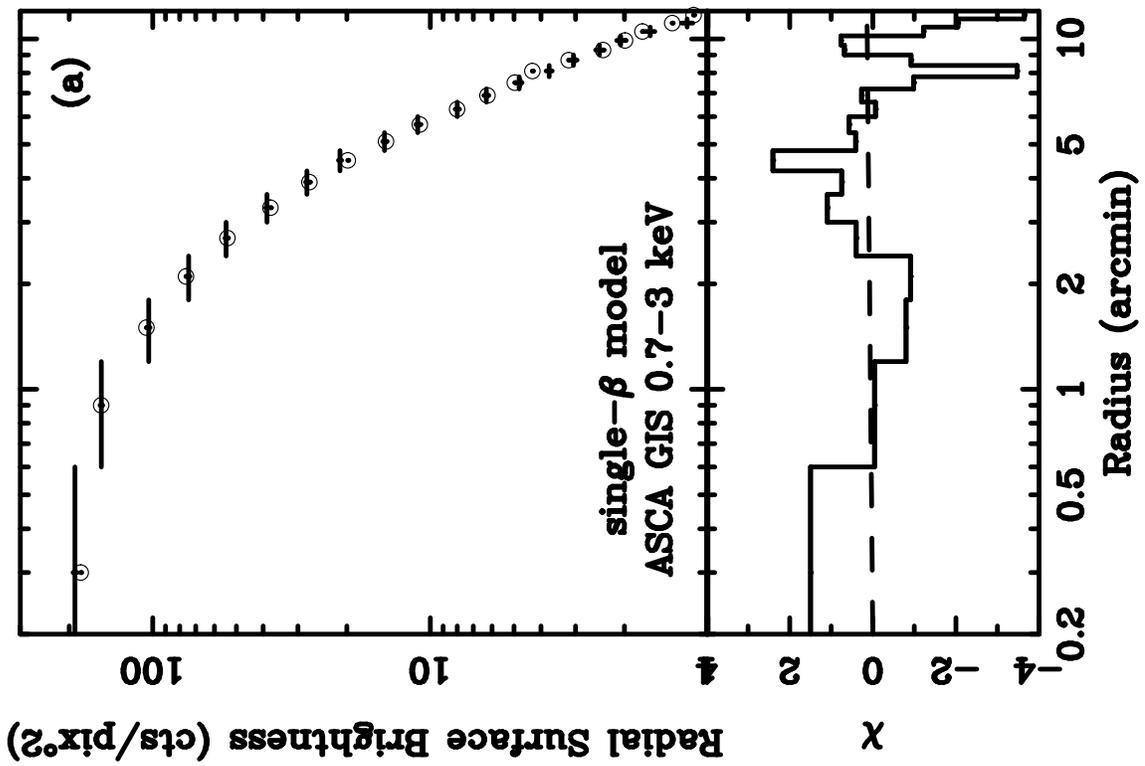
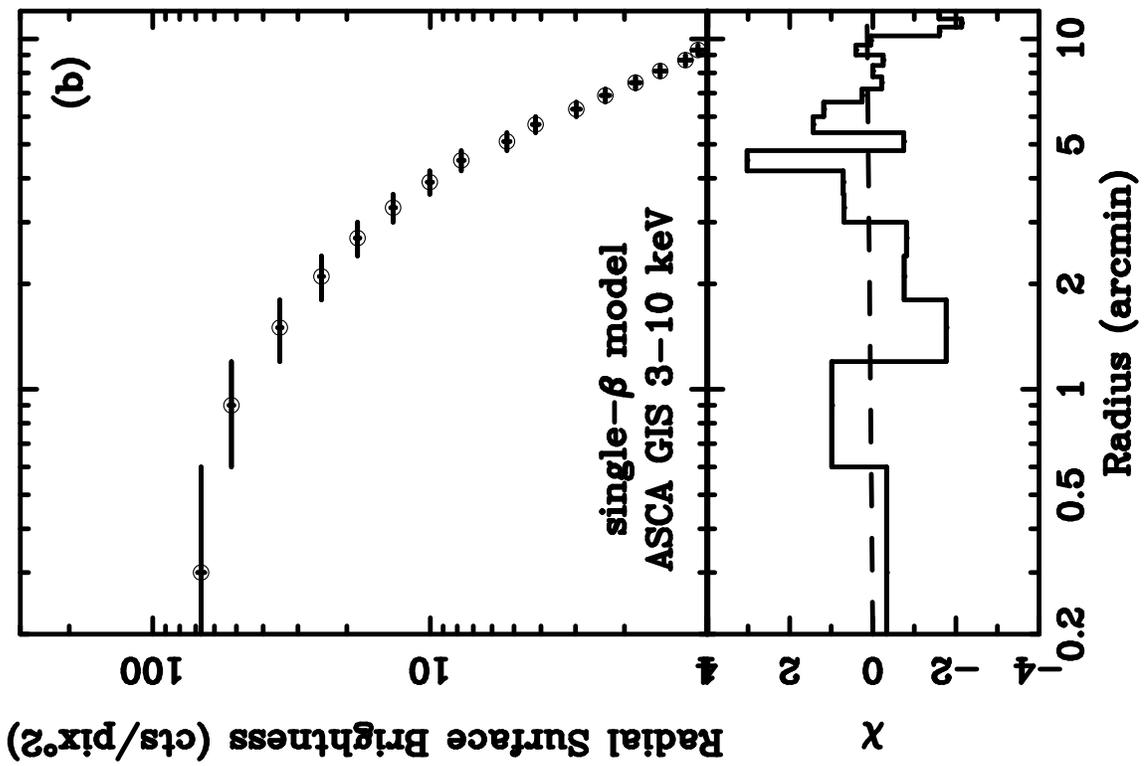
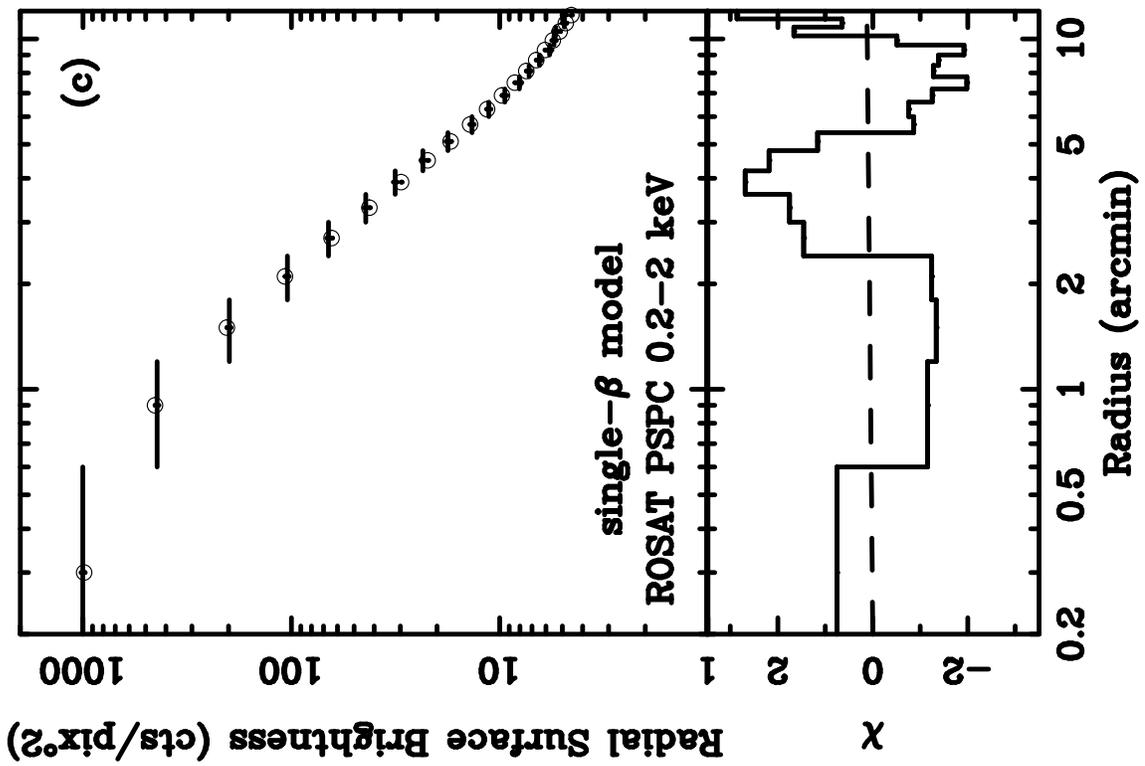
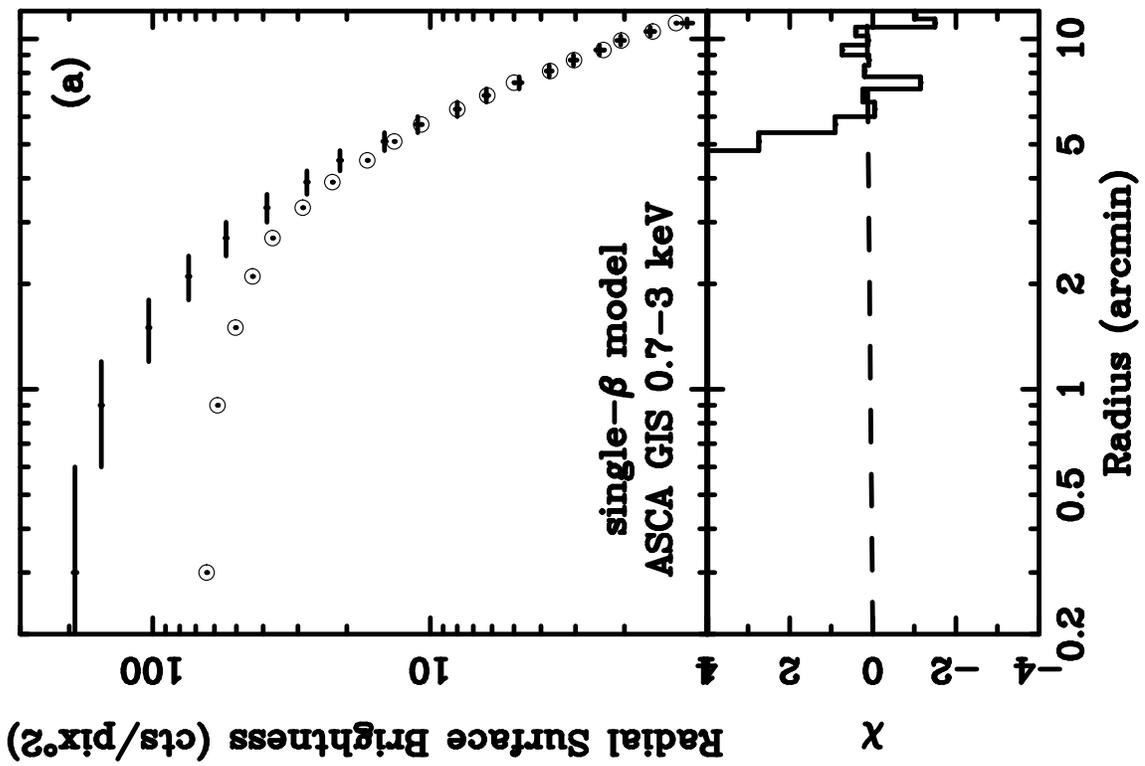
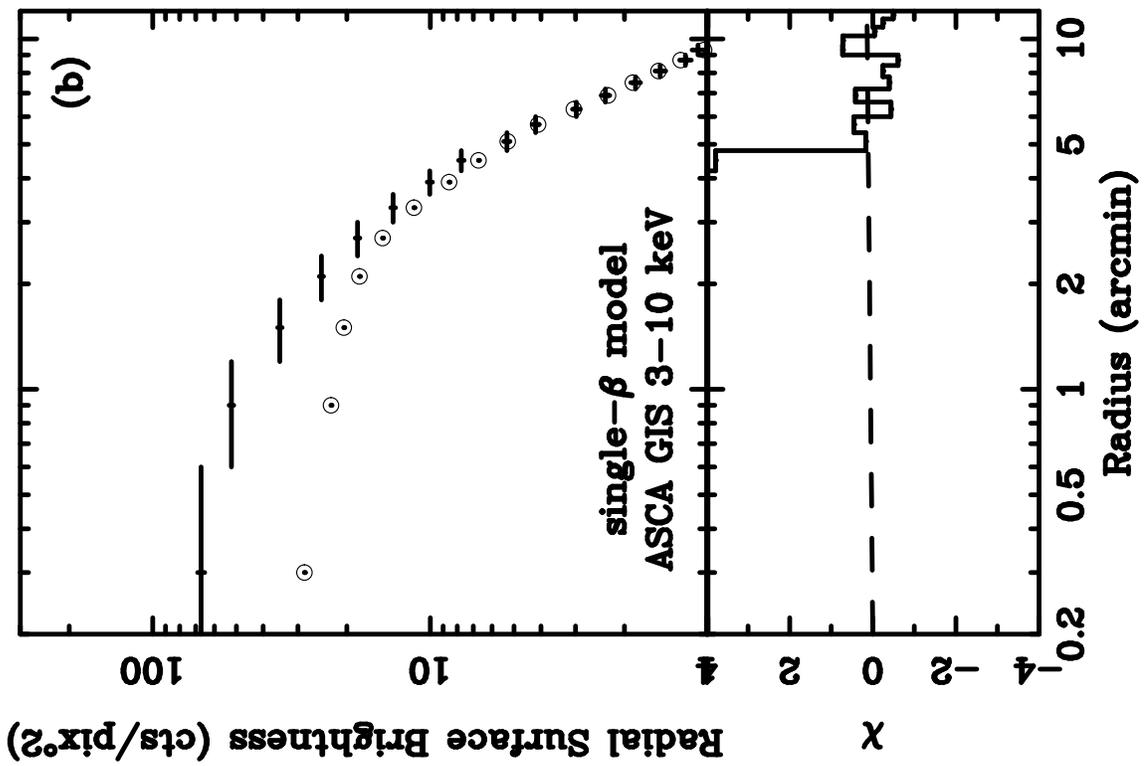
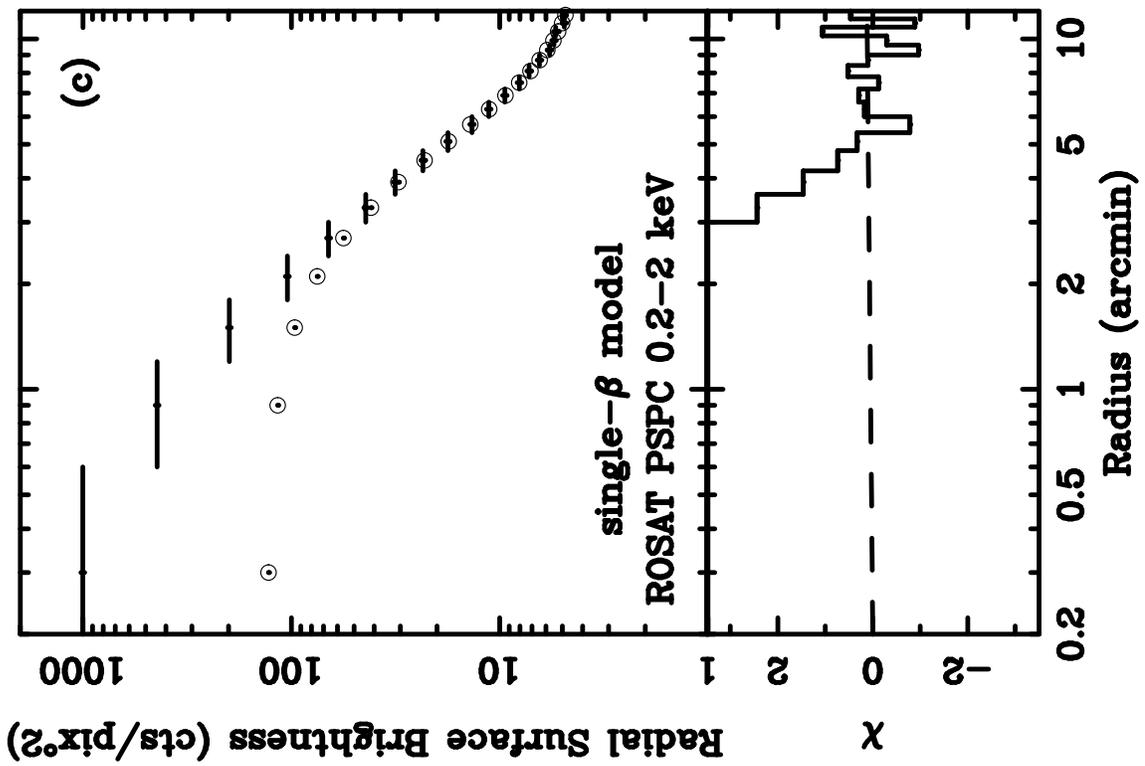
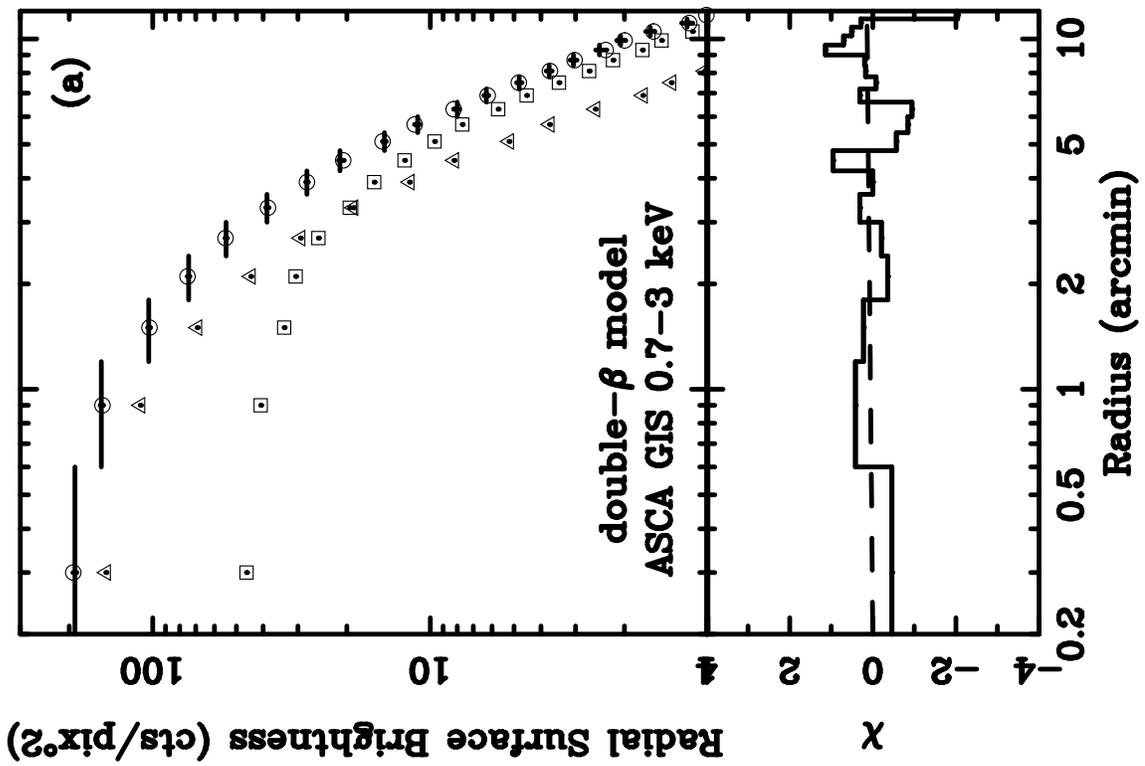
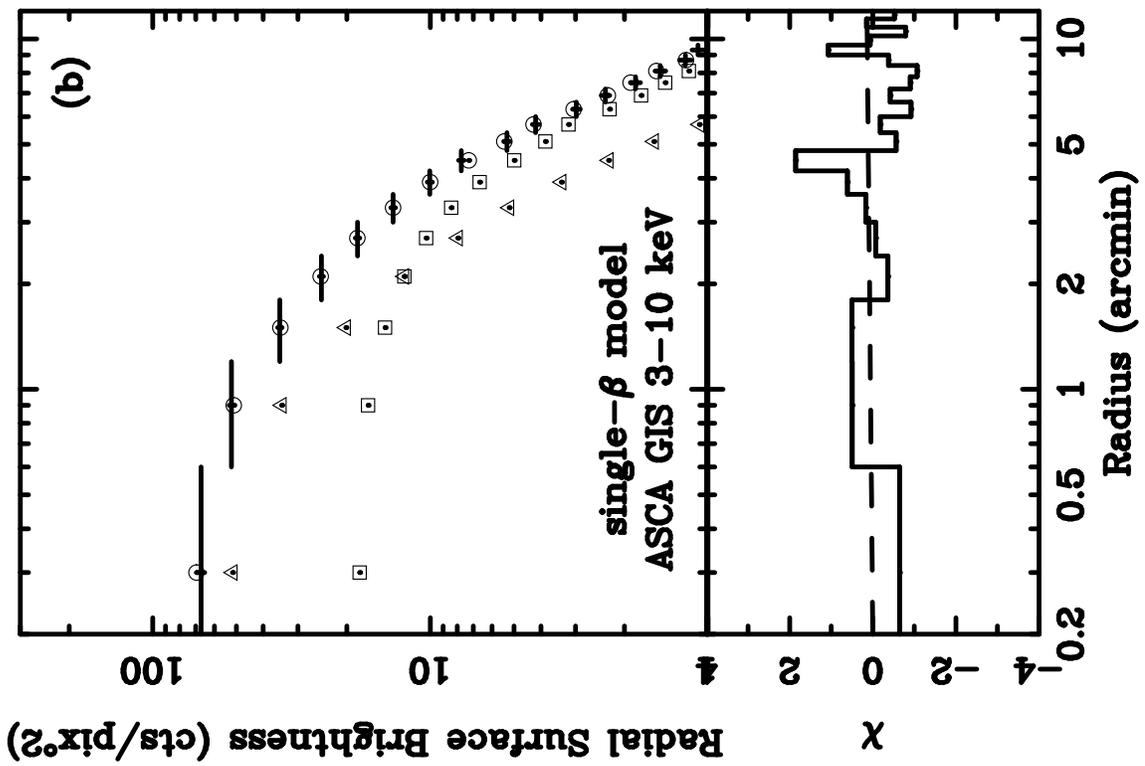
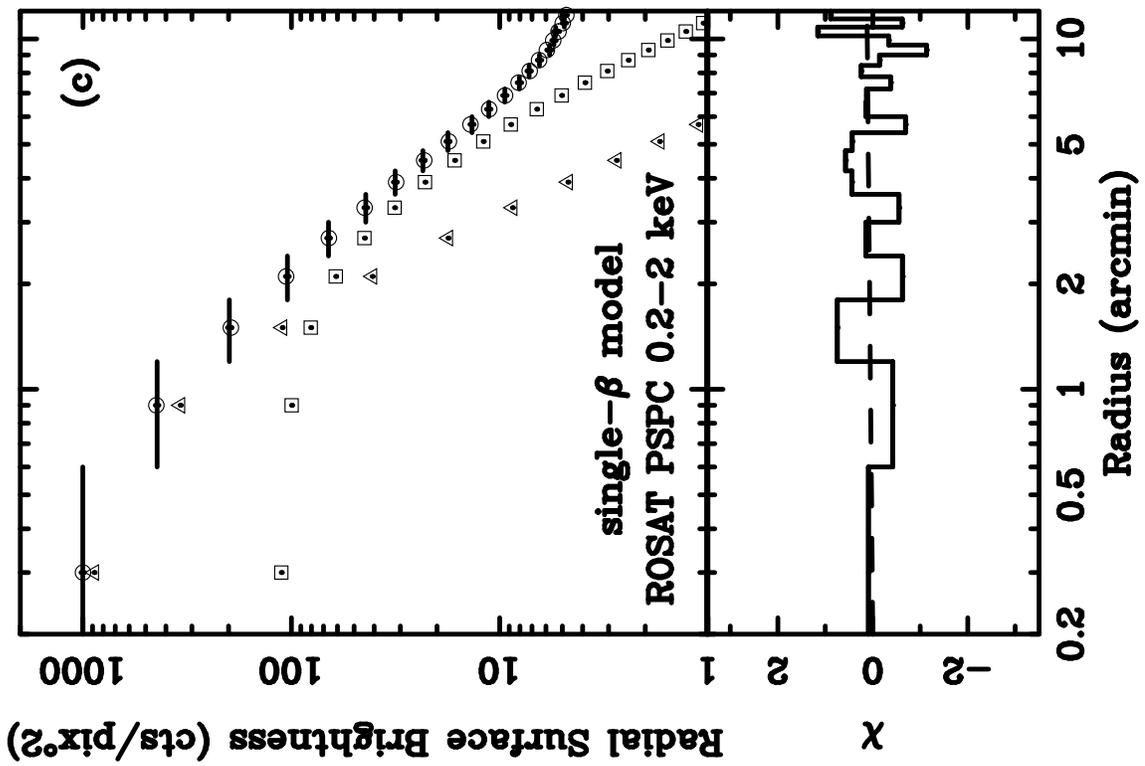
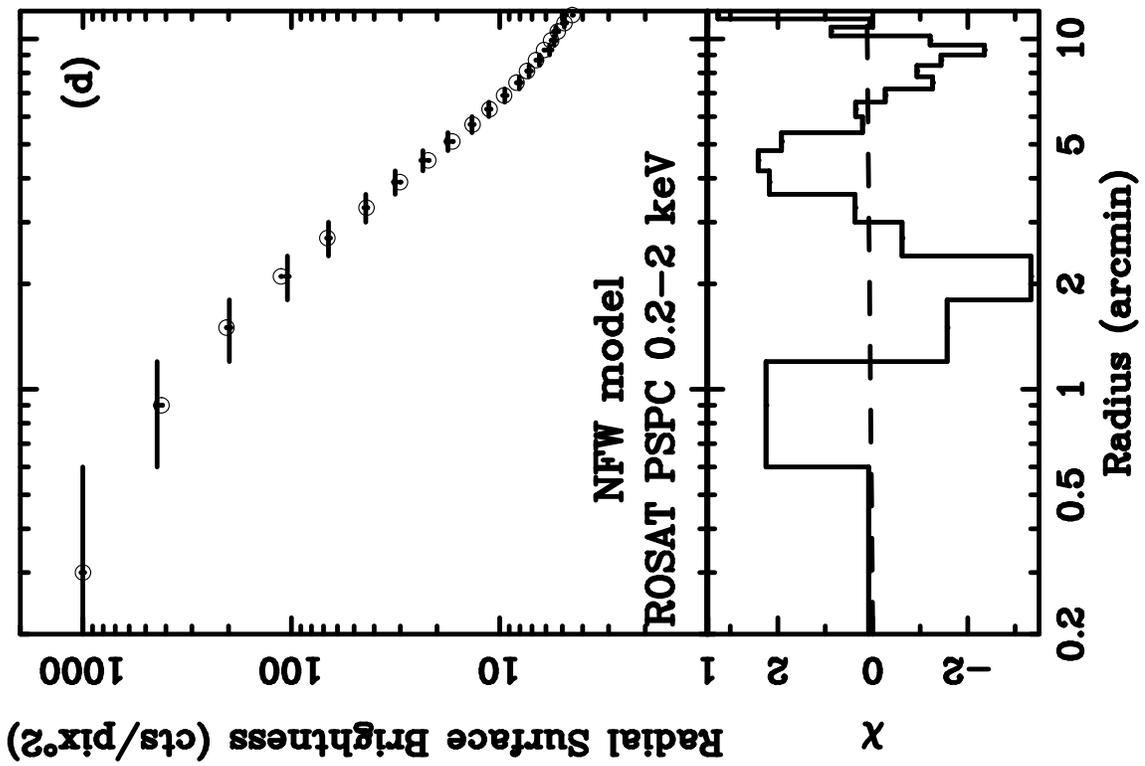
\end{document}